\documentclass{aastex631}
\usepackage{comment}
\usepackage{xcolor}
\usepackage{subfigure}

\newcommand{\nii}{[N \textsc{ii}]}
\newcommand{\sii}{[S \textsc{ii}]}
\newcommand{\oi}{[O \textsc{i}]}
\newcommand{\oiii}{[O \textsc{iii}]}
\newcommand{\ha}{H$\alpha$}
\newcommand{\hb}{H$\beta$}
\newcommand{\hd}{H$\delta$}
\newcommand{\galaxies}{77}
\newcommand{\extraplanar}{31}
\graphicspath{{./}{figures/}}

\begin{document}


\title{Shocked POststarburst Galaxy Survey. IV.  Outflows in Shocked Post-Starburst Galaxies Are Not Responsible For Quenching}

\author[0009-0007-2501-3931]{Antoniu Fodor}
\affiliation{Department of Physics \& Astronomy and Ritter Astrophysical Research Center, University of Toledo, Toledo, OH 43606, USA}
\author{Taylor Tomko}
\affiliation{Department of Physics \& Astronomy and Ritter Astrophysical Research Center, University of Toledo, Toledo, OH 43606, USA}
\author{Mary Braun}
\affiliation{Department of Physics \& Astronomy and Ritter Astrophysical Research Center, University of Toledo, Toledo, OH 43606, USA}
\author[0000-0001-7421-2944]{Anne M. Medling}
\affiliation{Department of Physics \& Astronomy and Ritter Astrophysical Research Center, University of Toledo, Toledo, OH 43606, USA}
\author{Thomas M. Johnson}
\affiliation{UKIRT Observatory, Institute for Astronomy, 640 N. A’ohoku Place, University Park, Hawaii 96720}
\affiliation{Department of Physics \& Astronomy and Ritter Astrophysical Research Center, University of Toledo, Toledo, OH 43606, USA}
\author{Alexander Thompson}
\affiliation{UKIRT Observatory, Institute for Astronomy, 640 N. A’ohoku Place, University Park, Hawaii 96720}
\affiliation{Department of Physics \& Astronomy and Ritter Astrophysical Research Center, University of Toledo, Toledo, OH 43606, USA}
\author{Victor D. Johnston}
\affiliation{Department of Physics \& Astronomy and Ritter Astrophysical Research Center, University of Toledo, Toledo, OH 43606, USA}
\author{Matthew Newhouse}
\affiliation{Department of Physics \& Astronomy and Ritter Astrophysical Research Center, University of Toledo, Toledo, OH 43606, USA}
\author[0000-0002-0696-6952]{Yuanze Luo}
\affiliation{William H. Miller III Department of Physics and Astronomy, Johns Hopkins University, Baltimore, MD 21218, USA}
\author[0000-0002-4235-7337]{K. Decker French}
\affiliation{Department of Astronomy, University of Illinois at Urbana-Champaign, 1002 W Green St, Urbana, IL 61801, USA}
\author[0000-0003-3191-9039]{Justin A. Otter}
\affiliation{William H. Miller III Department of Physics and Astronomy, Johns Hopkins University, Baltimore, MD 21218, USA}
\author[0000-0002-6582-4946]{Akshat Tripathi}
\affiliation{Department of Astronomy, University of Illinois at Urbana-Champaign, 1002 W Green St, Urbana, IL 61801, USA}
\author[0000-0003-1535-4277]{Margaret E. Verrico}
\affiliation{Department of Astronomy, University of Illinois at Urbana-Champaign, 1002 W Green St, Urbana, IL 61801, USA}
\author[0000-0002-4261-2326]{Katherine Alatalo}
\affiliation{Space Telescope Science Institute, 3700 San Martin Dr, Baltimore, MD 21218, USA}
\affiliation{William H. Miller III Department of Physics and Astronomy, Johns Hopkins University, Baltimore, MD 21218, USA}
\author[0000-0001-7883-8434]{Kate Rowlands}
\affiliation{AURA for ESA, Space Telescope Science Institute, 3700 San Martin Drive, Baltimore, MD 21218, USA}
\affiliation{William H. Miller III Department of Physics and Astronomy, Johns Hopkins University, Baltimore, MD 21218, USA}
\author[0000-0001-6670-6370]{Timothy Heckman}
\affiliation{William H. Miller III Department of Physics and Astronomy, Johns Hopkins University, Baltimore, MD 21218, USA}



\begin{abstract}
Shocked POst-starburst Galaxies (SPOGs) exhibit both emission lines suggestive of shock-heated gas and post-starburst-like stellar absorption, resulting in a unique subset for galaxy evolution studies. We have observed \galaxies{} galaxies that fulfilled the SPOGs criteria selection using the DeVeny Spectrograph on the Lowell Discovery Telescope. Our long-slit minor axis spectra detect \ha\ and \oiii\ in some SPOGs out to 6 kpc above the galactic plane. We find extraplanar ionized gas in \extraplanar{} targets of our sample overall. Using their internal and external kinematics, we argue that 22 galaxies host outflows with ionized gas masses ranging from $10^2 M_{\odot}$ to $10^5 M_{\odot}$. The rest are likely extended diffuse ionized gas. A positive correlation exists between AGN luminosity and the extraplanar gas extent, velocity dispersion, and mass---suggesting that the AGN may indeed drive the outflows detected in AGN hosts.
The low masses of the extraplanar gas suggest that these outflows are not depleting each galaxy's gas reserves.
The outflows, therefore, are not likely a significant quenching mechanism in these SPOGs.

\end{abstract}

\keywords{AGN, Quenching, Outflows, Galaxies, Post-starburst, SPOGs}


\section{Introduction} \label{sec:intro}

A bimodal distribution exists between blue spiral galaxies and red elliptical galaxies (\citealt{1958...Holmberg}, \citealt{1978...Larson..Tinsley...PecularNormal}, \citealt{Kauffmann...2003}, \citealt{2004...Baldry}). 
The two populations differ in color, morphology, and contents: the blue cloud tends to contain gas-rich spiral galaxies forming stars, while the red sequence often harbors gas-poor elliptical galaxies without much active star formation. 
Over the past 6 Gyr, the total mass of blue spiral galaxies has not increased (\citealt{2007....Noeske...bluespirals}, \citealt{2011...Wuyts....Bluespirals}) while the mass of red spirals has doubled (\citealt{2007...Bell...Red...Elliptical}, \citealt{2007...Faber...redellipticalgrowing}, \citealt{2012...Bell...Red}). Because of this, some portion of blue galaxies are believed to evolve and change shape, falling onto the red sequence after their star formation is suppressed. 
The physical process (or processes) that shut down, or ``quench", star formation in blue spirals is still under investigation.

In between bimodal distributions lies a ``green valley", much less populated than either the blue cloud or the red sequence.
The low density of objects in the green valley, coupled with the intermediate colors of those galaxies, suggest it might be an evolutionary stage in which some galaxies rapidly transition from blue to red by using up or losing their gas (\citealt{2007...Faber...redellipticalgrowing}, \citealt{2007...Salim...AGN...greenvalley}). 
\citealt{2014...Schawinski...pathways...green..valley} highlighted that many pathways could exist for blue spirals to evolve through the green valley: early-type galaxies thought to rapidly quench and late-type galaxies transitioning on a slower timeline.
Many late-type galaxies are found to be normal spiral galaxies that grew substantial amounts of intermediate and older stars and stay in the green valley for several Gyr. 

Star formation quenching occurs when gas in the interstellar medium is kept from condensing into stars by some physical mechanism(s).
This can happen as a result of either gas being removed from the galaxy or prevented from collapsing into stars.
Outflows (\citealt{2010...Feruglio..outflows}) or environmental effects like ram pressure stripping (\citealt{1972ApJ.....ram...pressure...gunn...gott}) may remove gas from the interstellar medium.
Similarly, feedback from stars or active galactic nuclei (AGN) can heat the gas and inhibit star formation (\citealt{2007...Kaviraj...stellarfeedback}, \citealt{2012...Hopkins...stellarfeedback}, \citealt{2016....Terrazas...agnfeedback}). 
Morphological quenching might also inhibit star formation once a galaxy reaches a certain size (\citealt{2013...Lillly...quenching}).
A combination of equally significant mechanisms acting on gas probably keeps a galaxy's star formation quenched (\citealt{2017...Smethurst....multiple...quenching}).
The main contributing mechanisms and what leads to star formation shutting off via negative feedback or expelled gas is still being studied.

Post-starburst galaxies (PSBs) have been an important class of galaxies used to study galaxy evolution.
PSBs are galaxies that had a recent starburst in the past Gyr and rapidly quenched their star formation since.
Classic post-starburst studies have focused on ``E+A" galaxies (\citealt{1996...Zabludoff...E+A}, \citealt{2004...Quintero....K+A}).
These are PSBs with high Balmer absorption (\hd), tracing the presence of a large intermediate-age A star population.
Additionally, galaxies with strong emission lines associated with star formation, such as \ha, are removed.
PSBs selected with such criteria, however, would be a limited sample to study when trying to understand quenching as a whole. 
Some rapid quenchers, like NGC 1266, would be excluded from typical E+A surveys because of its elevated emission lines from shock-heated gas present in the system (\citealt{2014...Alatalo...NGC1266}).
We must study galaxies that are earlier in their transition through the green valley to constrain what mechanisms cause successful long-term quenching.

The Shocked POst-starburst Galaxy Survey (SPOGS)\footnote{\dataset[https://www.spogs.org]{https://www.spogs.org}} aims to look for quenching signatures from a different perspective than traditional PSBs surveys.
Shocked post-starbursts (SPOGs) are still selected to have a dominant population of A stars; however, the survey also selects for PSBs with emission line ratios not exclusively tied to star formation (\citealt{2014...alatalo...spogs}, \citealt{Alatalo..SPOGs..2016A}).
By doing so, ongoing quenching mechanisms that might produce ionized gas are not excluded. 
This produces a unique subset of galaxies that is missed by other post-starburst surveys. 
SPOGs host a younger population of PSBs closer in time to their most recent starbursts and might catch galaxies in the process of shutting off their star formation (\citealt{Alatalo..SPOGs..2016A}, \citealt{2018...Decker...spogs...postburstage}).
SPOGs are chosen from their Sloan Digital Sky Survey (SDSS) nuclear data (\citealt{SDSS..DR7}) using the OSSY survey (\citealt{Oh...2011}, \citealt{2014...alatalo...spogs}, \citealt{Alatalo..SPOGs..2016A}).

Spatially resolved data is crucial in understanding quenching processes.
While ram pressure stripping might stop star formation on the edges of a galaxy, AGN feedback pushing gas out and heating it would start the quenching process in the galactic nucleus.
Evolving galaxies are believed to quench inside-out, with star formation first stopping by the galactic nucleus (\citealt{2019...Lin...inside-out...quenching}).
Studies like \citealt{Ho-2016-edgeon-outflows} leverage spatial data from integral field spectroscopy to study mechanisms that could affect galaxy evolution, like outflows and starbursts, inside and around galaxies.
Spatial information and emission line ratio diagrams like the BPT (\citealt{1981...BPT...NII}) or VO87 (\citealt{1987...VO87}) diagnostic diagrams can inform how significant a process like AGN feedback is throughout a galaxy.
Tracing the correlation between line ratio and gas velocity dispersion can further distinguish between extraplanar Diffuse Ionized Gas (eDIG) and shock-ionized gas, often associated with outflows \citep[e.g.][]{2007...Heald...edig...Vel.Disp...10-60, 2010ApJ...Rich...outflows...vel.disp, 2023...Johnston....3Dbpt}.

Spatially-resolved spectroscopy is needed to investigate SPOGs further as a PSBs catalog.
SPOGs show hints of large-scale winds from their nuclear fiber data from SDSS (\citealt{Alatalo..SPOGs..2016A}). 
The detection and frequency of those winds can help explain whether AGN or stellar feedback is suppressing star formation within these galaxies.
Work done by \citealt{Ho-2016-edgeon-outflows} has shown that normal edge-on blue spirals have outflows about 38\% of the time.
\citealt{2020...Roberts-Borsani...Manga...SF...outflow...population} detected outflows in ~20\% of star-forming galaxies in the MaNGA DR15 survey using \ion{Na}{1} D.
In comparison, \citealt{2020...Wylezalek....AGN...outflow...fraction...25percent} showed ionized gas outflows in around 25\% of their AGN MaNGA catalog.
We will explore whether SPOGs have elevated fractions of outflows and the drivers of possible outflows, such as AGN or star formation feedback.

This first paper is part of a series studying the properties of SPOGs galaxies and their place in the study of galactic evolution.
This paper studies the ionized emission along the minor axis of the observed SPOGs.
Following papers will look at star formation histories along the major axis that may be indicative of ongoing quenching.
In \textsection 2, we lay out the target galaxies observed using long-slit spectroscopy from the Lowell Discovery Telescope (hereafter referred to as LDT). 
In \textsection 3, we present extraplanar emission in observed SPOGs and discuss whether or not this gas is outflowing material.
In \textsection 4, we constrain the mass in the ionized outflows.
In \textsection 5, we discuss the implications of our findings on the SPOGs phase.

The cosmological parameters $H_0 = 70$ km s$^{-1}$, $\Omega_M = 0.3$, and $\Omega_\Lambda = 0.7$ are used throughout this paper. 

\section{Observations and Data Reduction} 

We obtain long-slit spectra along the minor axis of a sample of \galaxies{} shocked post-starburst galaxies to confirm the presence or absence of extraplanar ionized gas.

\subsection{Target selection}
\label{classification}
Observational targets were chosen from the SPOGs parent sample of 1067 galaxies in \citealt{Alatalo..SPOGs..2016A}.
Edge-on galaxies were preferred to easily identify outflowing material from the centers of the targets.
To ensure observed SPOGs were not strictly face-on and that any extraplanar gas would be visible above or below a galaxy, a cutoff of 0.7 was used for the ratio of the minor to major axis using images from the Sloan Digital Sky Survey (SDSS; \citealt{SDSS..DR7}).

Potential targets were then chosen to span a range of nuclear diagnostic classifications, according to the BPT/VO87 emission line diagnostic diagrams \citep{1981...BPT...NII,1987...VO87}.
These diagnostic diagrams compare line fluxes from pairs of nearby in-wavelength emission lines (\oiii\  $\lambda$5007 /\hb~vs. \nii\  $\lambda$6583 /\ha, \sii\  $\lambda\lambda$6716,31/\ha, and \oi\  $\lambda$ 6300/\ha) to characterize the hardness of the ionizing radiation field and define the main ionizing source (AGN, star-formation, or shock-heating).
In this work, classification requires a consensus majority of at least two out of the three emission-line diagnostic diagrams. 
Galaxies with different classifications for all three diagnostic diagrams were categorized based on their \oiii/\hb\ versus \sii/\ha\ nuclear classification.
The \sii\ / \ha\ line ratio diagnostic is emphasized because we are interested in separating ionized gas caused by shocks and AGN.
To be classified as an AGN-dominated SPOG, the BPT diagnostic diagram (\oiii/\hb\ versus \nii/\ha) must also classify the galaxy as an AGN.
Nuclear emission line data was taken from the SDSS, through the OSSY catalog (\citealt{SDSS..DR7}, \citealt{Oh...2011}).

\begin{figure}[ht]
   \centering
    
    \includegraphics[width=0.45\textwidth]{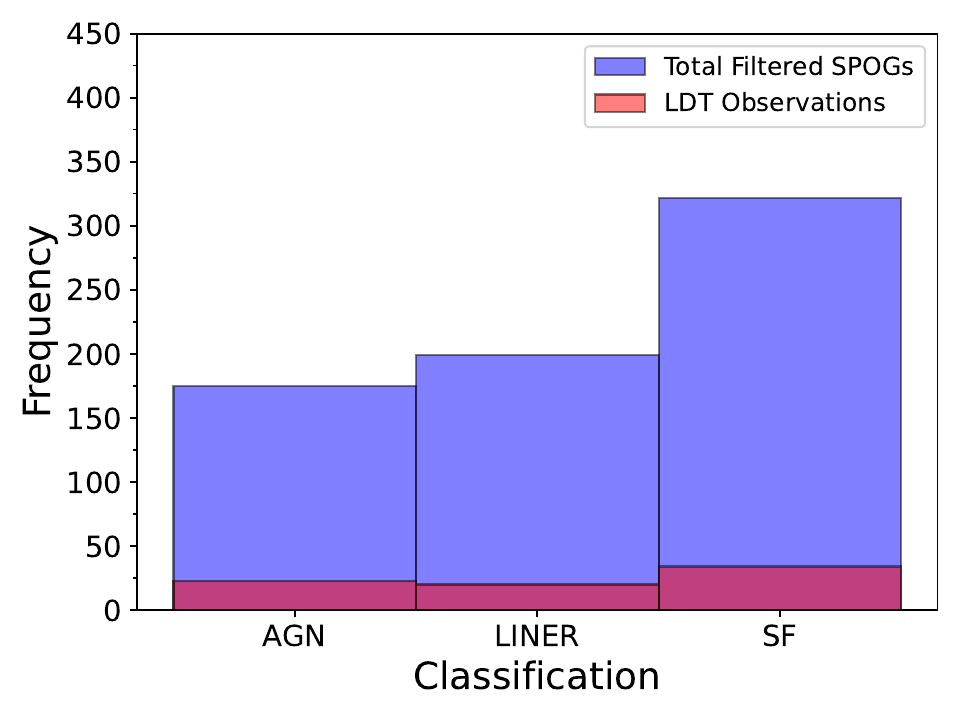}
    \hfill
    \includegraphics[width=0.45\textwidth]{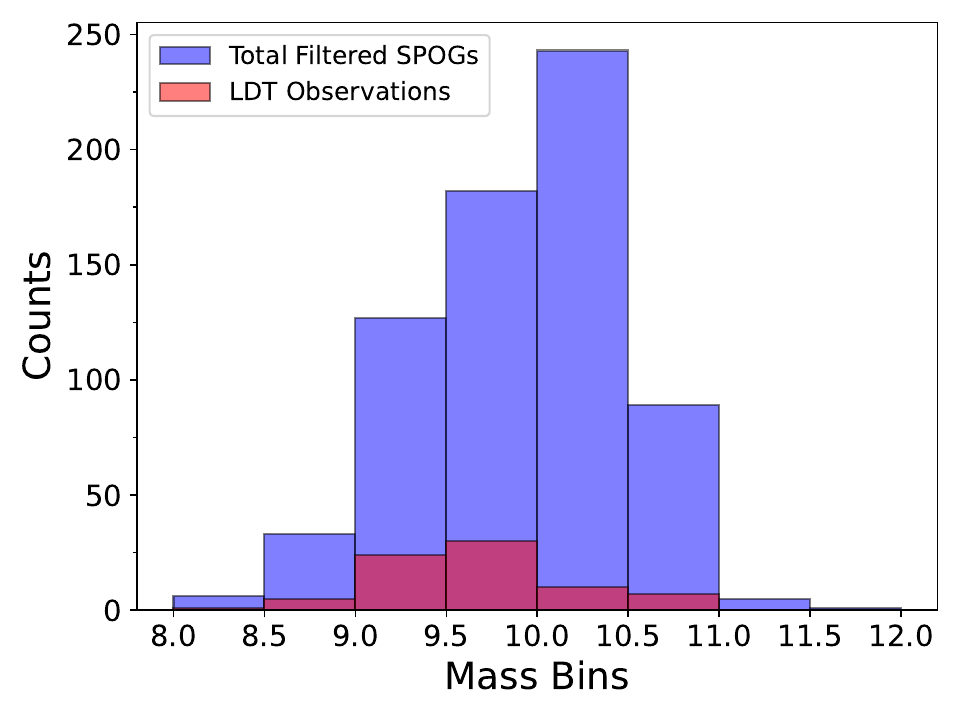}
    
    \caption{\textbf{Left:} Histogram comparing the observed and parent SPOGs samples. 
    They are separated in bins based on their emission line diagnostic classifications. 
    The parent sample is filtered to be within the same redshift range as the observed sample. \textbf{Right:} Histogram comparing the same observed and parent SPOGs samples divided into mass bins. 
    Masses are taken from the SPOGs Survey (Method outlined in \textsection 2.3 of \citealt{Alatalo..SPOGs..2016A}).
    }
\label{histogrampanels}
\end{figure}

Our goal is to search for triggers of quenching in the SPOGS sample as a whole. 
Figure \ref{histogrampanels} compares the observed sample from the LDT (red) with the overall SPOGs sample in the same redshift range (blue).
The left plot separates the two samples by classification and the right plot separates them by mass bins. 
The parent sample spans all three diagnostic classifications with a larger star-forming (SF) population than AGN or shock-heated (hereafter referred to as LINER-dominated gas).
The LDT observing campaign focused on collecting spectra of all three diagnostic classifications (AGN, LINER, SF) to not limit detections of extraplanar ionized gas that might be exclusive to a particular subset of SPOGs.
Table \ref{outflow_fraction_table} lists the number of SPOGs observed per classification in column 2.
The LDT sample spans a comparable range in stellar mass to the parent SPOGs sample, although our median stellar mass is slightly lower (9.8 vs 10).

The parent SPOGS sample covers a redshift z $<$ 0.2, but here we limit our sample to z $<$ 0.095 to keep the [\ion{S}{2}] from shifting out of our accessible spectral range.

\subsection{Observations}
We observed \galaxies{} SPOGs with the DeVeny spectrograph at the 4.3-m Lowell Discovery Telescope near Flagstaff, Arizona \citep{deveny-instrument}. 
Observations were carried out between August 2019 and April 2023 over 33 nights.
 
The slit is 2.5' long and was set to a width of 1.15".
The DeVeny spectrograph was used in combination with a 500 g/mm grating (blazed at $\sim$5500 \AA).
The GG420 rear order-blocking filter was also installed for these observations. 
These settings allow observations that span 4500 to 7600 \AA\ with a spectral resolution of R$\sim$1500. 
This wavelength range gives us the following lines needed for our analysis: \oiii, \hb,  \nii, \ha, and \sii. 
We observed long-slit position angles aligned with both the major and minor axes of each galaxy.
Due to the length of the long-slit, we can fit any SPOG fully within the slit along either axis.
In this paper, we only make use of the minor axis slit positions.
Figure \ref{2d_spectral_emission}'s first column has SDSS images of three SPOGs targets observed with DeVeny's long-slit, for reference.
The blue rectangle represents the long-slit's width and positioning on each galaxy to gather minor axis exposures.

Each galaxy's exposure ranged from 600 seconds to 1200 seconds depending on the g-band apparent magnitude of the galaxy, its redshift (to account for worse sensitivity at redder wavelengths), and nightly conditions. 
Typically, galaxies were observed for 900 seconds of integration time per exposure.
A list of all the galaxies observed and their observing dates are provided in Appendix \ref{Observing Appendix}.
Although conditions varied, this typically resulted in limiting surface brightness sensitivities of $1-8 \times 10^{-16}$ erg\,s\,cm$^{-2}$\,arcsec$^{-2}$.
The spatial pixel scale of our 2D spectra is 0.34 arcsec per pixel, sampling seeing that typically ranged from 1-3 arcsec.
At least three exposures were taken for the perpendicular minor axis of a galaxy to improve signal through coadding.
Most exposures include small random dithers courtesy of the telescope pointing system, which were corrected for when coadding.
Calibration frames (bias, dome flats, and Cadmium-Argon-Mercury lamp spectra for wavelength calibration purposes) were also taken at the beginning of each night.
For flux calibration, standard stars that had an angular separation $<$ 30 degrees were observed.

\subsection{Data Reduction and Emission Line Analysis}
\begin{figure*}
\includegraphics[scale=0.55, trim=0cm 1.6cm 0cm 1.5cm,clip]{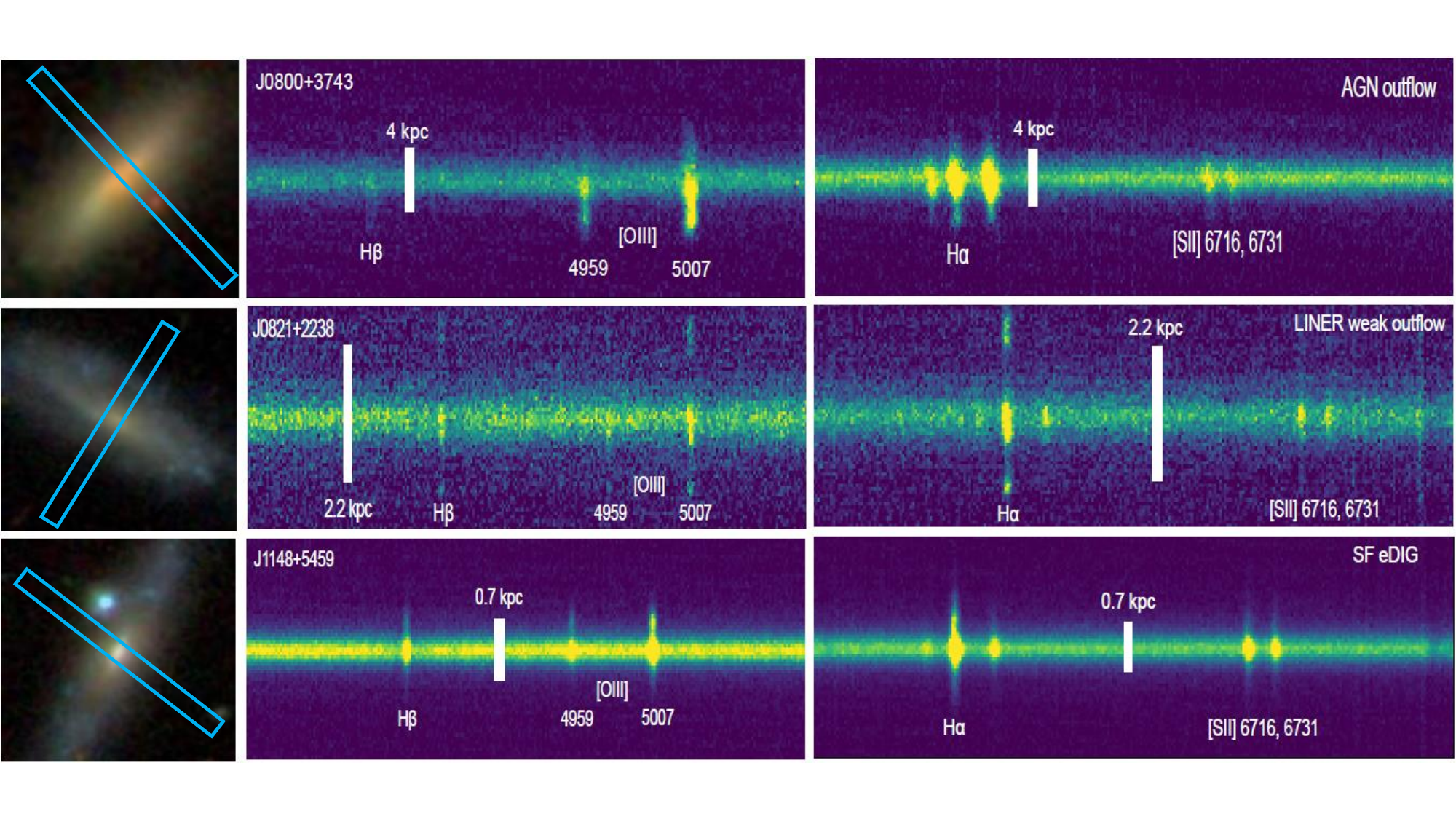} \\
\caption{Long-slit spectroscopy of three different SPOG galaxies observed with the LDT.
The slit was placed along the minor axis of each edge-on target. 
White bars along the 2D spectra mark the radial extent we define as internal to the galaxy using a 1D Gaussian fit to the continuum.
We define ionized gas external to the white bar's limits as extraplanar ionized gas.
This observing technique is capable of confirming and characterizing galactic winds in post-starburst target galaxies along the minor axis.
}
\label{2d_spectral_emission}
\end{figure*}

With the absence of a reduction package commonly used for the DeVeny spectrograph at the time of our observations, we developed a data reduction pipeline in python to reduce our 2D spectral data. 
The python reduction follows the basic steps of data reduction outlined in section 3 of \citealt{MasseySpectroscopy}.
To summarize the pipeline's steps, we start by subtracting a median-combined bias frame from each galaxy and standard star frame.
Then, we align frames of the same object and median-combine to produce a single galaxy or star file. 
A median-combined flat field is then divided from each galaxy and star file.
The median of the sky continuum above/below the spectra is subtracted per column across each 2D spectra.
We use the arc lamp frames taken during individual observing nights to wavelength calibrate each galaxy and star spectra.
Each galaxy spectra is then flux calibrated with an observed standard star.
We corrected for galactic extinction from the Milky Way (\citealt{2011...Schlafy...Redoing...Gal...Extinction}, \citealt{1998...old...Schlegel...galactic...extinction}). 
Figure \ref{2d_spectral_emission} shows example reduced 2D long-slit spectral data for 3 SPOGs.

\begin{figure*}
  \centering
  \subfigure{\includegraphics[width=0.45\textwidth, trim=0cm 0cm 0cm 0cm,clip]{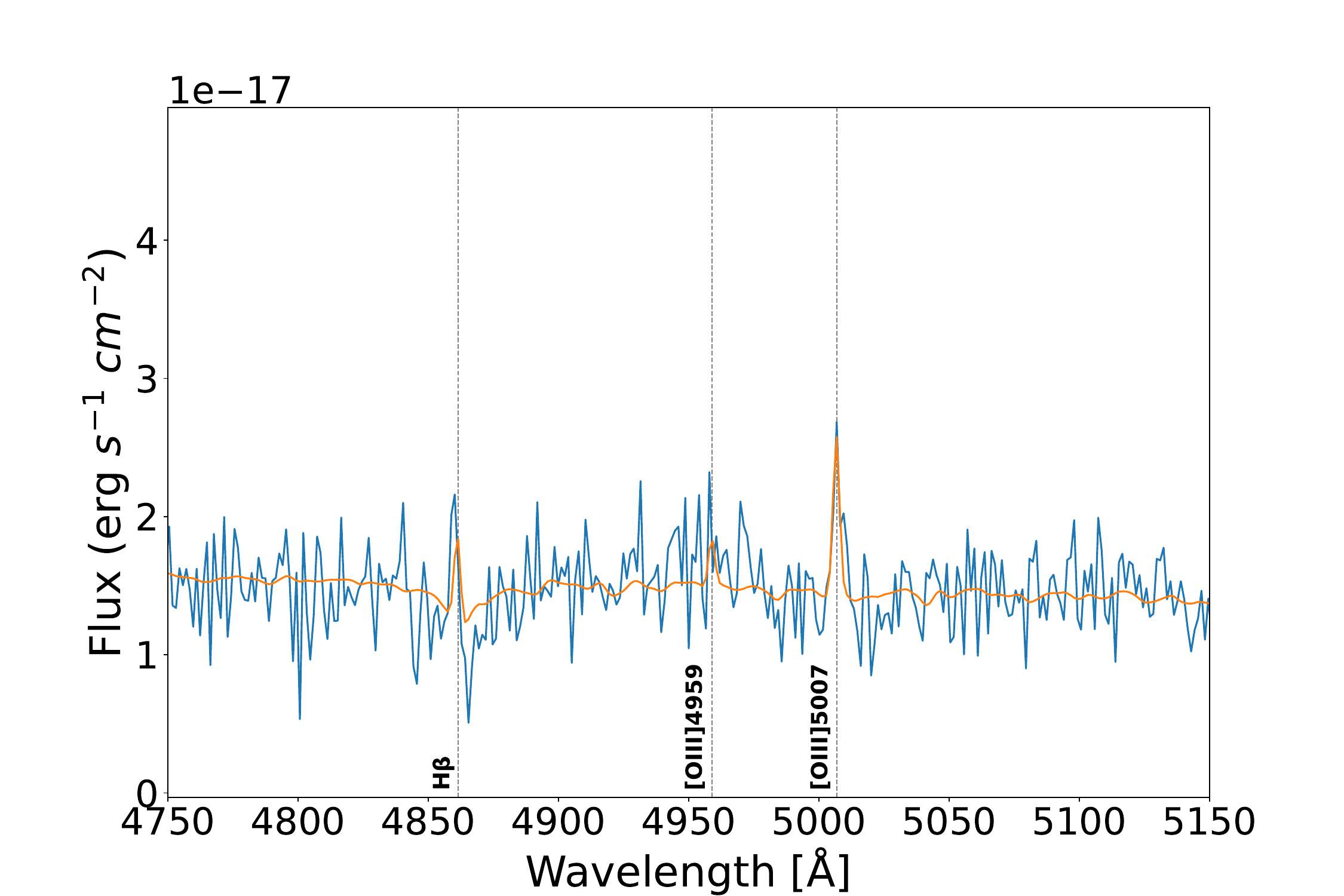}}
  \subfigure{\includegraphics[width=0.45\textwidth, trim=0cm 0cm 0cm 0cm,clip]{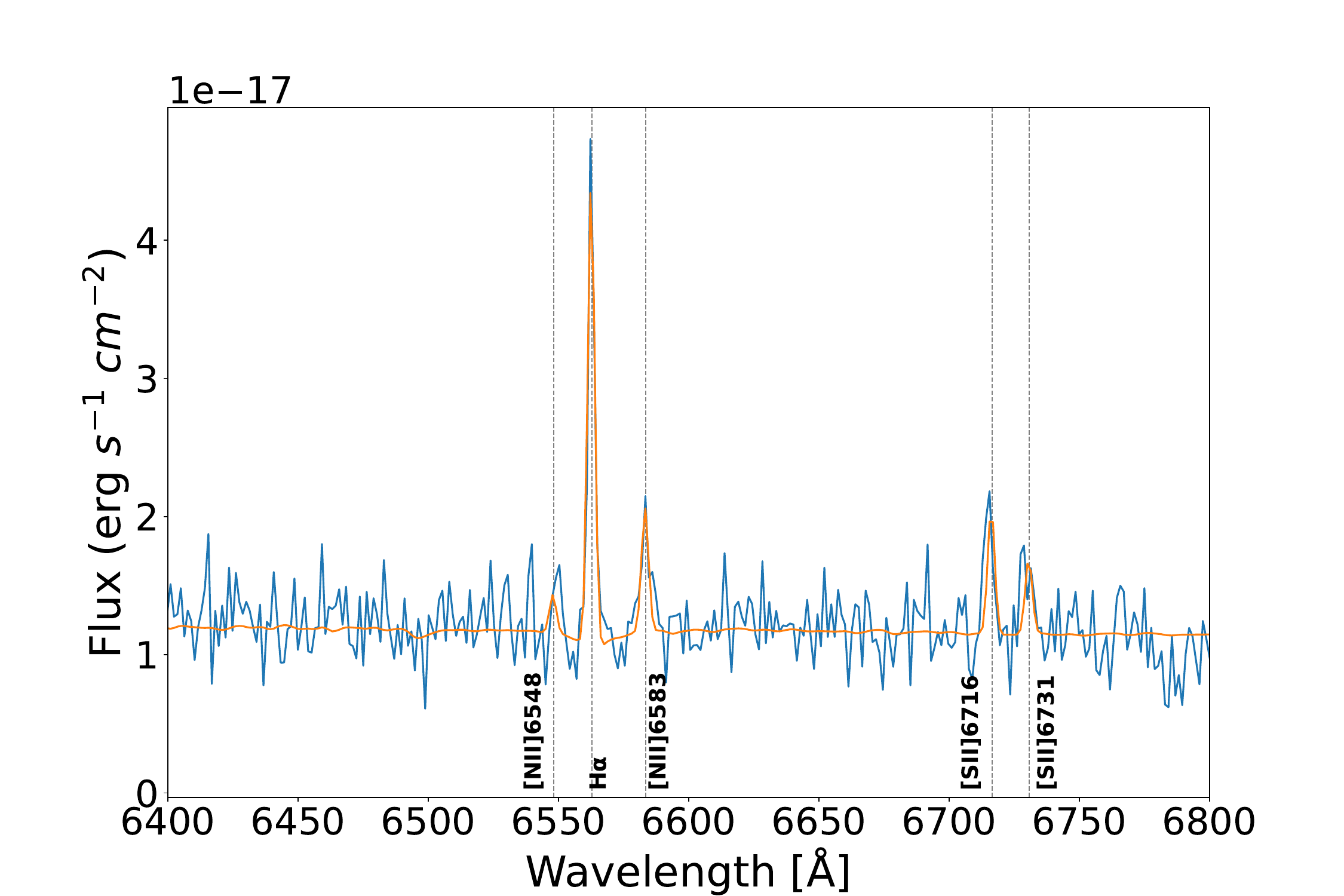}}
  \subfigure{\includegraphics[width=0.9\textwidth, trim=0cm 0cm 0cm 0.3cm,clip]{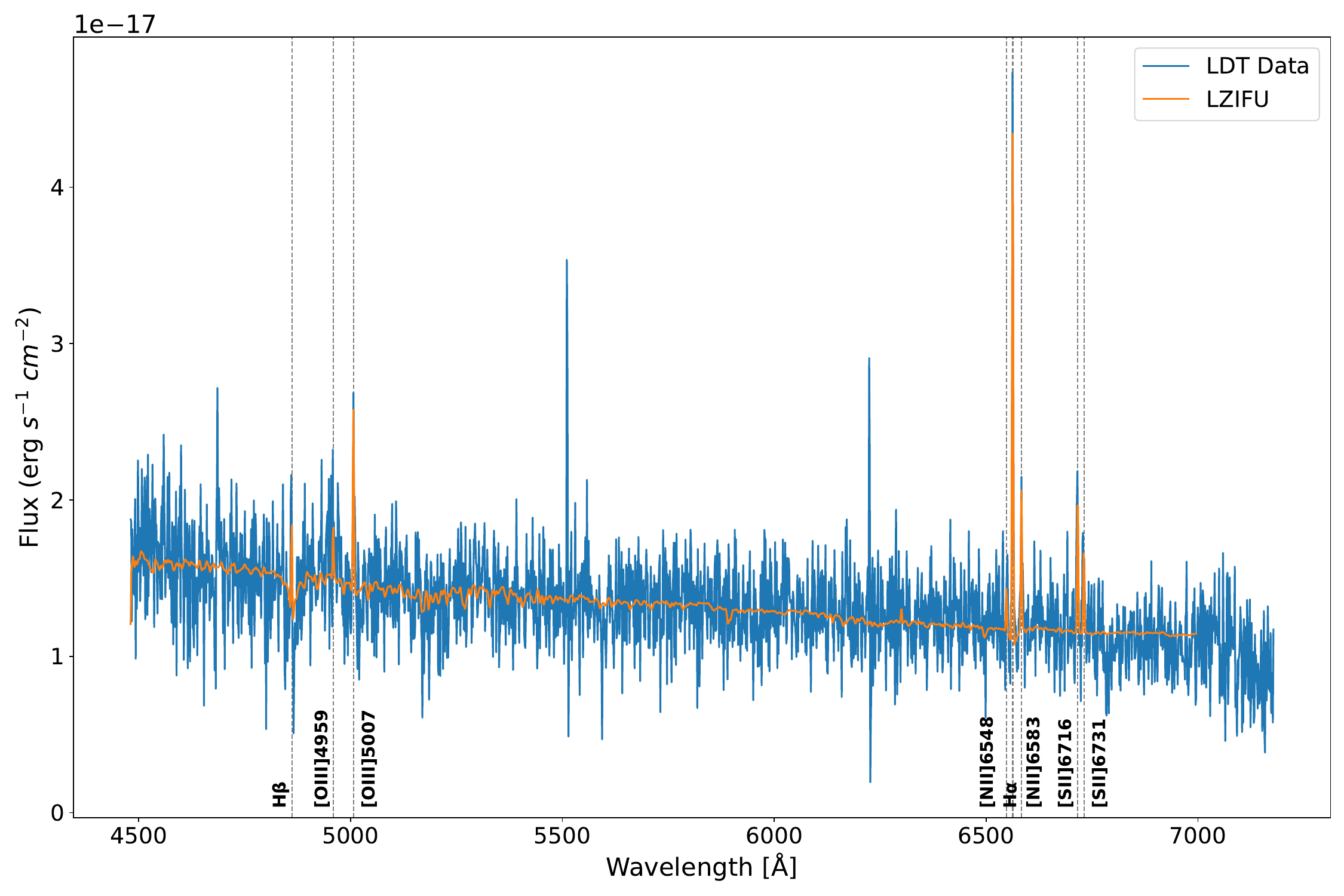}}
  \caption{
  The central binned spectral row of the galaxy J0821+2238 (blue) with LZIFU’s fit overlaid in orange.
  \textbf{Top Left:} Zoom-in on the bluer wavelengths of the spectrum alongside the LZIFU fit. 
  \textbf{Top Right:} Same as top left, but now focused on the red end. 
  \textbf{Bottom panel:} The full galactic spectrum. 
    The emission lines relevant to this work (\hb, \oiii $\lambda\lambda$ 4959,5007, \ha, \nii $\lambda\lambda$ 6548,6583, and \sii $\lambda\lambda$ 6716,6731) are labeled and marked with gray dashed lines. 
}
  \label{lzifu_ldt_fit}
\end{figure*}

We reshaped the 2D data into a 3D dataset to run through the 3D line fitting tool, LZIFU (\citealt{LZIFU...Ho...2016}).
LZIFU is an IDL-based package designed to fit emission lines and produce kinematic maps for 3D integral field spectroscopy data cubes.
It performs the Penalized PiXel-Fitting (pPXF) (\citealt{2004PASP..116..138C}, \citealt{ppxf2016}) routine to fit stellar absorption using the SSPPadova spectral templates (\citealt{2005...Delgado...SSP}). 
SSPPadova uses a theoretical model of stellar population evolution following Padova isochrones (for more detail, see \citealt{LZIFU...Ho...2016}, \citealt{2005...Delgado...SSP}, and \citealt{2005MNRAS...Martins...SSPPadova}).
The SSPPadova library uses stellar populations that range in age from 4 to 18 Gyr with 0.5 and 1 Z$_{\sun}$.
Then, LZIFU fits the emission lines using the Levenberg-Marquardt least-square method. 
We use it to fit the \hb, \oiii, \ha, \nii, and \sii\ emission lines present in our optical spectra.
Figure \ref{lzifu_ldt_fit} shows the spectrum at the center spatial pixel of one SPOG, J0821+2238, and the LZIFU fit overplotted.

For some galaxies, ionized gas is present when visually inspecting the 2D spectra but does not meet a signal-to-noise threshold of 3 to classify as detected emission. 
In such cases, rows of the 2D long-slit data are binned together to improve signal-to-noise levels.
To retain spatial information given the small size of SPOGs galaxies, we create 2-, 3-, 4-, and 5-pixel binned 2D spectra to also be analyzed with LZIFU.

\section{Detecting Extraplanar Emission} 
\label{detection}
Extraplanar ionized gas is detected using the spatial dimension of our long-slit spectroscopy. 
We look for ionized gas in any of the emission lines used for diagnostics (\ha, \nii, \sii, \hb, \oiii).
Detected extraplanar emission needs to extend past the galaxy's edges.
To define what is internal and external to a specific galaxy in our sample, we fit a one-component Gaussian fit to the stacked and median-combined continuum of that individual 2D spectrum.
Any data points outside of two standard deviations from the center of the fitted Gaussian are considered extraplanar.
Figure \ref{2d_spectral_emission}'s middle and right columns show 2D spectral slices for three galaxies. 
The middle column shows a wavelength range around the \oiii\ emission line and the right column for \ha. 
In each column, a white bar is shown that marks the extent of each galaxy's \textit{internal} continuum using our Gaussian fitting method. 
Ionized gas emission above our signal-to-noise threshold of 3 outside of a galaxy's continuum is classified as extraplanar ionized emission.

The circumgalactic medium (CGM) has a dynamic relationship between a galaxy and its halo, hosting gas flowing out of a galaxy's ISM and gas falling back onto the ISM (\citealt{2023...cgm...review}).
As part of the CGM, diffuse ionized gas (eDIG) extends out past the limit of the galactic disk generally to distances of a few kpc (\citealt{1973...Reynolds....electrondensity...edig}, \citealt{1996...edig...extent}, \citealt{Haffner2009}, \citealt{2019...Levy...eDIG}).
From a study of 22 edge-on spiral galaxies, eDIG seems to extend out to similar scale heights to neutral gas in the CGM (\citealt{2023....eDIG....cgm...neutralgas...Lu}).
Kinematic information is key in identifying extraplanar gas that is entrained in outflows or that might be eDIG.
Many works, including that of \citealt{2019...Levy...eDIG}, find that eDIG is warm ionized gas surrounding galaxies that lags behind the rotational speed of the galactic disc. 
The ionization mechanism for eDIG is thought to be from star formation in the disk.
This diffuse gas is ionized but has low ranges of velocity dispersions (e.g., \citealt{2007...Heald...edig...Vel.Disp...10-60}, \citealt{2019ApJ...edig...Vel.Disp...20-60...Boettcher}, \citealt{2019...Levy...eDIG}).
Extraplanar gas could also be the result of outflows from the galaxy pushing gas into the surrounding medium.
This can be caused by physical mechanisms, like a central AGN, violently pushing gas out of a galaxy.
Outflows can be identified by elevated kinematics in the \ha\ and \oiii\ emission (\citealt{woo2016}, \citealt{Ho-2016-edgeon-outflows}, and \citealt{concas2019}).

In \textsection \ref{common extraplanar fraction}, we discuss the fraction of SPOGs that host outflows and their characteristics. 
In \textsection \ref{kinematics and outflows? eDIG?}, we describe the kinematics of the extraplanar gas and if gas is truly outflowing or extraplanar diffused ionized gas.
In \textsection \ref{extent extraplanar}, we measure how far beyond the galaxies the detected extended emission reaches.
In \textsection \ref{what is ionizing}, we discuss the primary ionization sources of these outflows using line diagnostic diagrams. 

\subsection{How Common is Extraplanar Ionized Gas?}    
\label{common extraplanar fraction}
\begin{figure}[ht]
    \centering
    
    \includegraphics[width=0.6\textwidth]{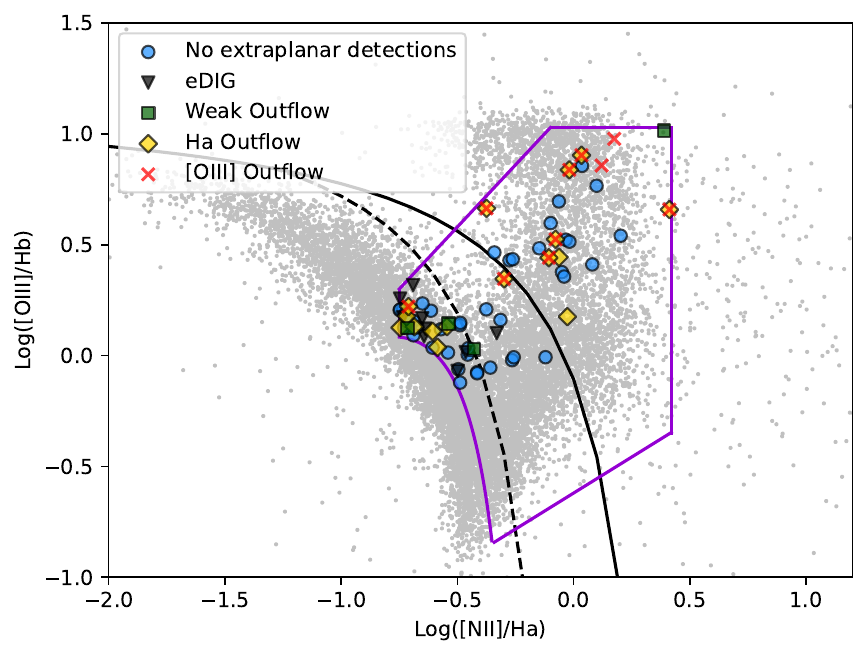}
    \hfill
    \caption{BPT emission line diagnostic diagram for \oiii/\hb\ versus \nii/\ha. 
    Blue points are LDT-observed SPOGs without confirmed extraplanar gas.
    Black triangles denote SPOGs with extraplanar gas detected with low velocity dispersions (see \textsection \ref{kinematics and outflows? eDIG?}).
    Green squares are SPOGs with a weak outflow from their velocity dispersions and extraplanar detections in either \ha\ or \oiii. 
    Yellow diamonds mark LDT observed SPOGs with confirmed extraplanar \ha\ ionized gas and velocity dispersions indicative of an outflow. 
    Red crosses mark observed SPOGs with \oiii\ extraplanar ionized gas we label as outflowing.
    The gray points are observations from the SDSS (\citealt{2012...SDSS...DR9}).  
    The purple line marks the boundary in which SPOGs fall on the BPT according to shock models (for a further discussion of the shock models and SPOGS criteria, see \citealt{Alatalo..SPOGs..2016A}.) 
    The star formation boundary is denoted by the black solid curve (\citealt{2006...Kewley...BPT...SFcurve}) and the dashed curve (\citealt{Kauffmann...2003}).
    While \oiii\ outflow detections typically fall in the elevated line ratio region of the BPT, \ha\ and weak outflows are detected in SPOGs anywhere within the shock boundary.
    }
\label{extraplanar_bpt_full_sample}
\end{figure}

Out of the \galaxies{} galaxies in our sample, \extraplanar{} show evidence of extraplanar emission.
Table \ref{outflow_fraction_table} summarizes these results over the total LDT sample, per our diagnostic classification (\textsection \ref{classification}), and per mass bins.
We find extraplanar detections of \ha\ in 29 galaxies and of \oiii\ in 15 galaxies.
Figure \ref{extraplanar_bpt_full_sample} shows our full sample of observed SPOGs plotted on a BPT diagram based on their nuclear emission lines from OSSY (\citealt{Oh...2011}).
The SDSS DR9 full sample is added as background gray points, for reference (\citealt{2012...SDSS...DR9}). 
LDT SPOGs observations are denoted as to whether extraplanar ionized gas is detected or not.
Extraplanar ionized gas is marked based on how elevated its external velocity dispersion is compared to its internal velocity dispersion (eDIG, weakly outflowing, or outflowing gas). 
A further discussion of the kinematics of the extraplanar gas will be in \textsection \ref{kinematics and outflows? eDIG?}.
Extraplanar gas detections span a range of ionization properties on the BPT diagram, but trends are seen with \oiii\ outflowing gas and low velocity dispersion gas (eDIG)---generally at opposite ends of the BPT. 

We separate the observed sample into diagnostic classifications (\textsection \ref{classification}) to see how the extraplanar emission fraction changes between populations. 
Both AGN and star-forming SPOGs host extraplanar \ha\ emission in 44\% of each classification.
These two classifications differ when it comes to \oiii\ extraplanar emission, where AGN SPOGs make up most of the \oiii\ extraplanar gas detections (10 out of 15 galaxies). 
AGN SPOGs have an extraplanar \oiii\ fraction of 44\% while SF galaxies only have extraplanar \oiii\ detections in 12\% of their sample.  
LINER SPOGs, in contrast to both SF and AGN, have lower H$\alpha$ and \oiii\ extraplanar detection rates at 20\% and 5\%, respectively.
For the total sample, \ha\ is detected more often than \oiii\ with gas detection fractions of 37.7\% and 19.5\%, respectively.

SPOGs with extraplanar gas span most of the mass range of our subsample.
Our sample is skewed towards lower mass galaxies compared to the parent sample of SPOGs (see the right panel of Figure~\ref{histogrampanels}). 
There is no significant trend in detection rate, with most detections lying between 9 and 10.5 log($M_*$/$M_{\odot}$).
\citealt{Ho-2016-edgeon-outflows} found a similar lack of a trend between galactic mass and extraplanar fraction when looking at a sample of 40 wind-dominated galaxies in the Sydney-AAO Multi-object Integral field spectrograph Galaxy Survey (SAMI). 

Of note, we rarely detect extraplanar \hb\ ionized gas. 
When we do, \hb\ generally does not extend as far as \oiii\ detections do. 
In all but one case (for a LINER SPOG), \hb\ is not detected if \oiii\ is not also present. 
For that one galaxy, J1156+2829, it is possible that the recent starburst drove gas out to cause the detected outflow, causing ionized gas detected with \hb\ and not \oiii.

\begin{table*}
\begin{tabular}{cccccccc}
\hline\hline
& Count & Tot. Extraplanar & Tot. Extraplanar & H$\alpha$ & H$\alpha$  & [\ion{O}{3}] & [\ion{O}{3}] \\
&  & Gas Detections & Fraction & Count & Fraction & Count & Fraction\\
(1) & (2) & (3) & (4) & (5) & (6) & (7) & (8) \\ 
\hline
Total & \galaxies & \extraplanar & 40.3\% & 29 & 37.7\% & 15 & 19.5\% \\
AGN & 23 & 12 & 52.2\% & 10 & 43.5\% & 10 & 43.5\% \\
LINER & 20 & 4 & 20\% & 4 & 20\% & 1 & 5\% \\
SF & 34 & 15 & 44.1\% & 15 & 44.1\% & 4 & 11.8\% \\
\hline
Masses &  &  &  &  &  & &  \\
($\log(M_*)$) &  &  &  &  &  & &  \\
8-9 & 6 & 2 & 33.3\% & 2 & 33.3\% & 1 & 16.6\% \\
9-9.5 & 24 & 10 & 41.7\% & 10 & 41.7\% & 3 & 12.5\% \\
9.5-10 & 30 & 14 & 46.7\% & 12 & 40\% & 8 & 26.7\% \\
10-10.5 & 10 & 4 & 40\% & 4 & 40\% & 2 & 20\% \\
10.5-11 & 7 & 1 & 14.3\% & 1 & 14.3\% & 1 & 14.3\% \\
\hline\hline
\end{tabular}
\caption{Confirmed extraplanar ionized gas in the LDT observed SPOGs sample. 
Column 4 details the total extraplanar gas detection fraction for both \ha\ and \oiii, with columns 6 and 8 showing detection fractions for \ha\ and \oiii\ individually. 
These extraplanar fractions are given for the total SPOGs population, per BPT classification, and per mass bins.}
\label{outflow_fraction_table}
\end{table*}

\subsection{Velocity Dispersion and Outflow/eDIG classification}
\label{kinematics and outflows? eDIG?}
One important question about the detected extraplanar ionized gas is whether or not this gas is outflowing material from the center of each SPOGs galaxy, as the nuclear fiber data from SDSS hinted at in \citealt{Alatalo..SPOGs..2016A} for the overall SPOGs population.
Our survey may be detecting extraplanar diffuse ionized gas \citep[eDIG:][]{2019ApJ...edig...Vel.Disp...20-60...Boettcher, 2007...Heald...edig...Vel.Disp...10-60, 2000...Rossa...edig...2kpc} instead.
In order to confirm extraplanar ionized gas as an outflow, we need to look at the kinematic information of the relevant emission lines.

Traditionally, outflows are characterized by fitting multiple Gaussian components to emission lines like \oiii\ or \ha\ for broad and narrow regions (e.g., \citealt{2020ApJ...Liu...outflows}). 
LZIFU does allow us to fit multiple Gaussians to emission lines, but we chose to not do this given the low signal-to-noise of our extraplanar emission line detection. 
However, velocity dispersion of the ionized gas is a good signpost as to outflow status. 
Studies of outflowing extraplanar gas show an increasing velocity dispersion from the center of a galaxy outward \citep[e.g.][]{2010ApJ...Rich...outflows...vel.disp, 2011...Vel.Disp....Rich...elevated.line.ratios}.
Outflowing gas typically has high extraplanar velocity dispersion typically above 100 km s$^{-1}$ (\citealt{2010ApJ...Rich...outflows...vel.disp}).
In contrast, eDIG has lower velocity dispersion values \citep[10-60 km s$^{-1}$;][]{2019ApJ...edig...Vel.Disp...20-60...Boettcher,2007...Heald...edig...Vel.Disp...10-60}. 
While the LDT's velocity resolution is 200 km s$^{-1}$ per the spectrograph's spectral resolution, LZIFU's single component fit is able to resolve kinematic features from emission lines below an instrument's limits when the signal-to-noise ratio is high enough by removing the instrumental velocity dispersion from the model fits (see \citealt{LZIFU...Ho...2016}).
If LZIFU were fitting spuriously narrow lines (by fitting noise peaks), the lowest velocity dispersion points would also have the lowest signal-to-noise.
We investigated our fits and found no evidence that this was happening at our signal-to-noise threshold above velocity dispersions of at least 30 km s$^{-1}$.

Thus, the difference between internal velocity dispersion and the external velocity dispersion can be a crucial category by which to argue whether or not gas detections are related to outflow or eDIG. 
Table \ref{mass_galaxy_property_table} has columns for the average internal velocity dispersion and the average extraplanar velocity dispersion per galaxy.
If the internal average dispersion is higher than the external and the extraplanar velocity dispersion is low, then that argues in favor of the extraplanar gas being eDIG or some similar non-outflow related ionized gas.
In contrast, if the internal average dispersion is lower than the extraplanar gas (or at a similar high velocity dispersion), then that argues in favor of the extraplanar ionized gas being part of a driven outflow.

For our work, we will label three categories for extraplanar gas kinematic status: 
\begin{itemize}
\item eDIG - if the external velocity dispersion is less than 60 km s$^{-1}$ 
\item Outflows - if the external velocity dispersion is at least one sigma higher than the internal velocity dispersion
\item Weak outflows - if the external velocity dispersion is higher than 60 km s$^{-1}$, but less than one sigma above the internal velocity dispersion
\end{itemize}

In our sample, nine galaxies (eight SF and one LINER) have average internal and external velocity dispersions $<$ 60 km s$^{-1}$. 
We classify their extraplanar gas as possibly eDIG and not part of a galactic outflow; these galaxies are marked in Table~\ref{mass_galaxy_property_table} with a dagger. 
Four galaxies have an average external and internal velocity dispersion that meet our criteria of weak outflows, three of which fall in the SF SPOGs classification and one in the AGN classification.
The majority of extraplanar gas detections (11 AGN SPOGs, 4 SF SPOGs, and 3 LINER SPOGs) have higher external velocity dispersions than the internal velocity dispersions of their host galaxies.
This gas is likely to be entrained in outflows.

\subsection{How Far Does the Extraplanar Emission Extend?}    
\label{extent extraplanar}

For SPOGs with extraplanar gas confirmation, the spatial extent of the detection is an important characteristic. 
We define the furthest extent as the distance between the center of the galaxy through to the furthest data pixel above our signal-to-noise threshold, converted to kpc.
We note that this is a \textit{detected} furthest extent, as extraplanar gas may exist even further out but below our detection limit.
In table \ref{mass_galaxy_property_table}, we list the furthest detectable extent of extraplanar \ha\ (Column 2) and \oiii\ (Column 3) gas for each galaxy.

\begin{figure*}
  \centering
  \subfigure{\includegraphics[width=0.75\textwidth, height=0.25\textheight, trim=0cm 0cm 0cm 0cm,clip]{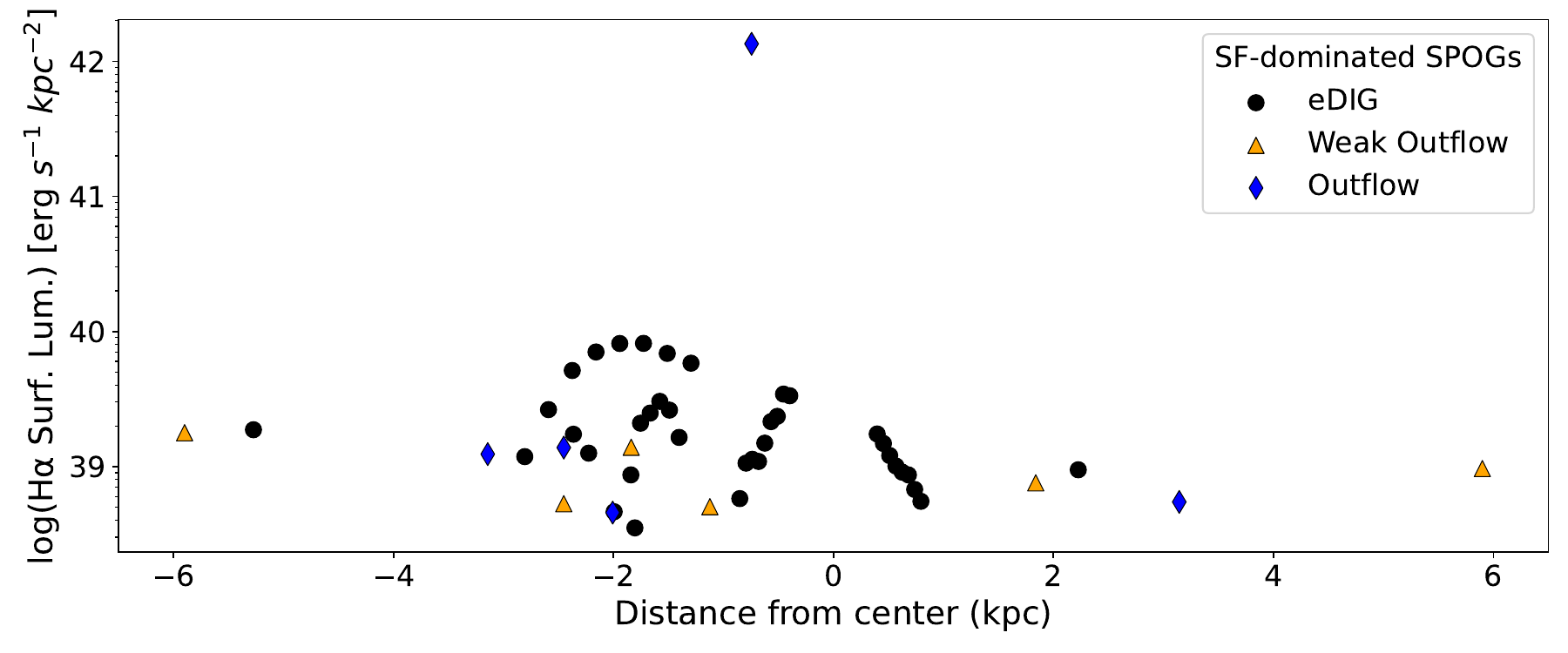}}
  \subfigure{\includegraphics[width=0.75\textwidth, height=0.25\textheight, trim=0cm 0cm 0cm 0cm,clip]{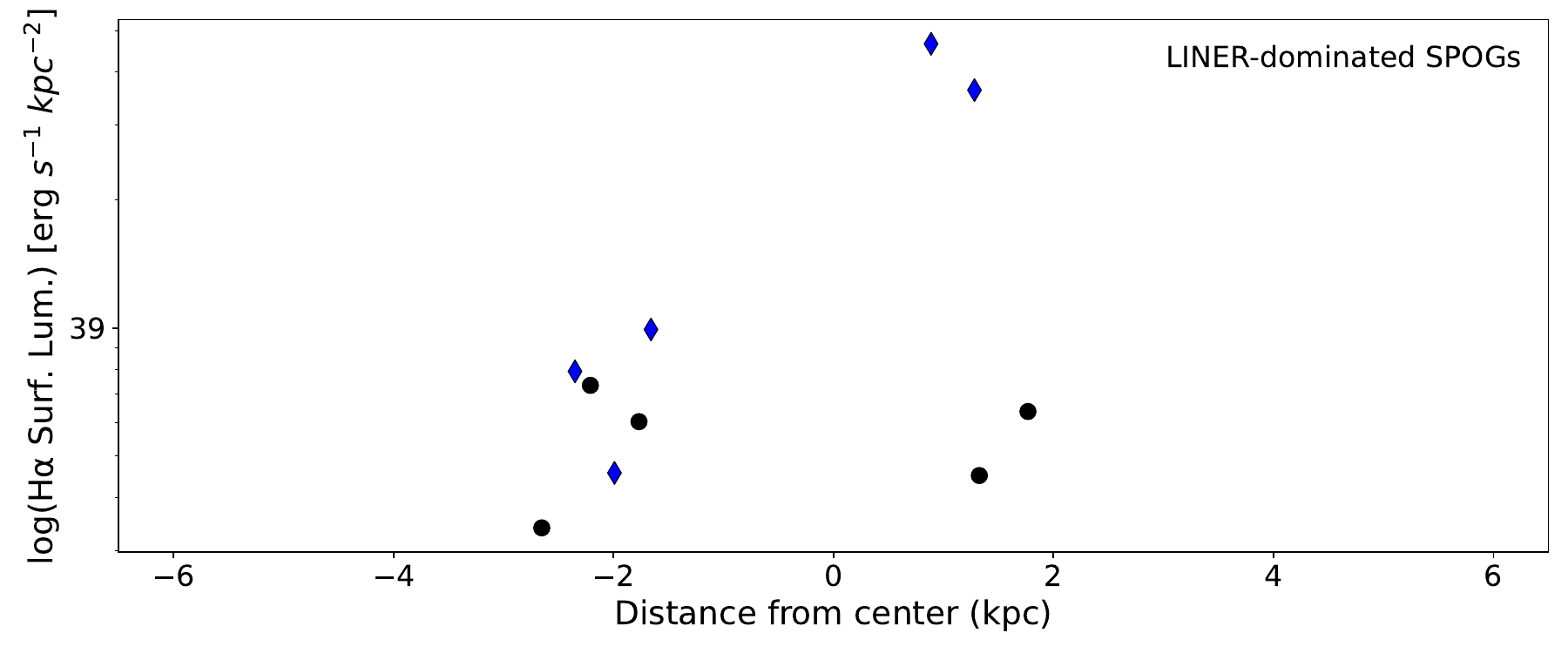}}
  \subfigure{\includegraphics[width=0.75\textwidth, height=0.25\textheight, trim=0cm 0cm 0cm 0cm,clip]{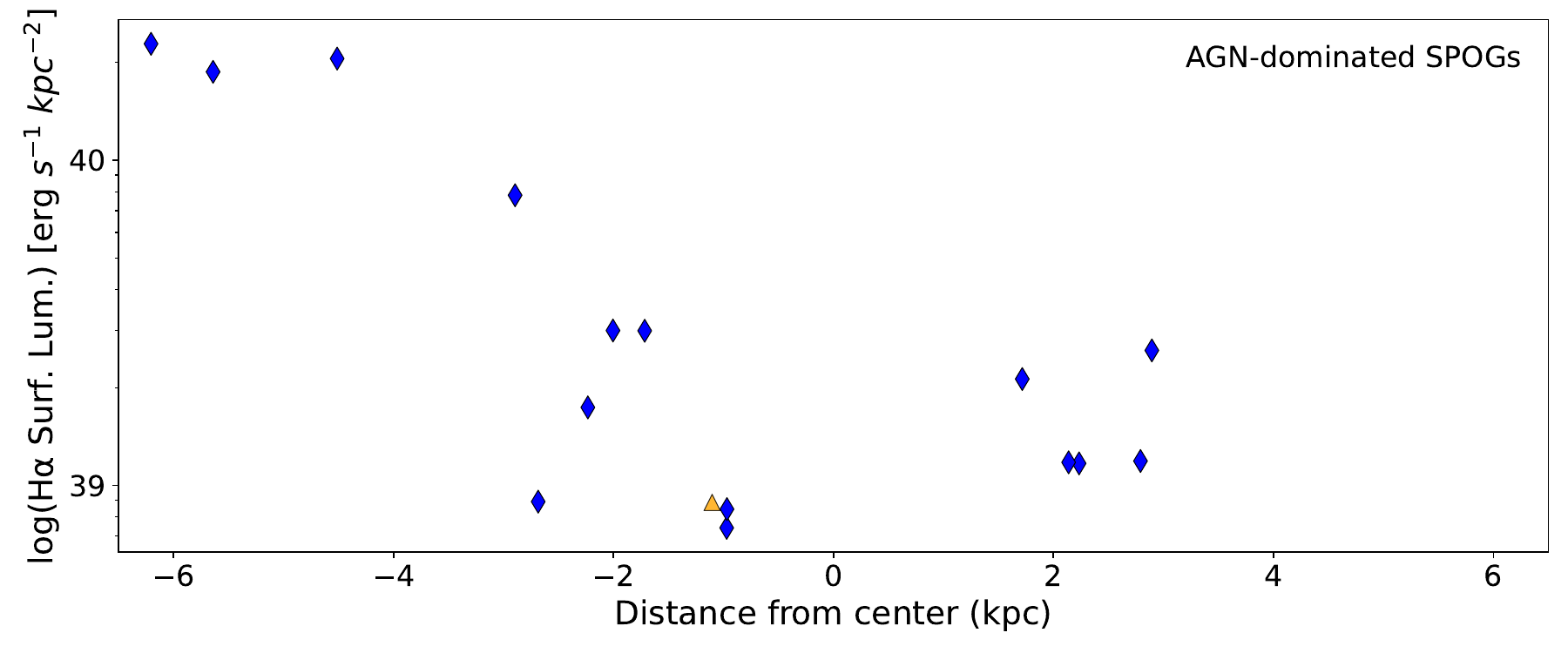}}
  \caption{Detected extraplanar ionized gas in H$\alpha$. 
H$\alpha$ surface luminosity density is in units of erg s$^{-1}$ kpc$^{-2}$.
SPOGs are divided by the emission line diagnostic classification scheme (\textsection \ref{classification}).
\ha\ emission within each galaxy is masked out. 
Points are colored and shaped based on their kinematic classification (\textsection \ref{kinematics and outflows? eDIG?}).
Black circles represent ionized gas associated with eDIG.
Orange triangles label gas with velocity dispersions suggesting weak hints of outflows.
Blue diamonds mark gas with clear outflow-related velocity dispersion.
All three classifications have detections out to 2 kpc, but SF and AGN SPOGs have ionized gas that extends out to 5-6 kpc away from the galactic center.
}
\label{halpha_extraplanar}
\end{figure*}

The maximum extent of our extraplanar emission does not depend on nuclear classification.
Figure \ref{halpha_extraplanar} shows \ha\ luminosity for extraplanar gas plotted as a function of distance from the center of each galaxy. 
SPOGs are separated based on their nuclear classification into SF, LINER, and AGN plots in the top, middle, and bottom panels, respectively.
Internal \ha\ gas detections within the galactic continuum (as earlier defined using a Gaussian-sigma threshold) are masked out.
Points are shaped and colored based on their kinematic classification as outflows or eDIG. 
While three galaxies show detectable \ha\ emission extending past 5 kpc (two SF galaxies and one AGN), 
the overall sample typically shows extraplanar ionized gas detections out to around 2 kpc from respective galactic centers.
\ha, on average, extends out further for SF SPOGs than for AGN and LINER SPOGs.
The average furthest extent for \ha\ is 2.3 kpc for AGN, 2.1 kpc for LINER, and 2.5 kpc for SF SPOGs.
In contrast, \oiii\ detections have their furthest distances for AGN SPOGs.
The average furthest extent for \oiii\ is 2.4 kpc for AGN, 2.2 kpc for LINER, and 2 kpc for SF SPOGs.

The scale of extraplanar gas in Table \ref{mass_galaxy_property_table} varies from barely outside the limits of the galaxy (0.7 kpc) to a significant distance past the edges (6.2 kpc). 
Previous ionized gas outflow studies have shown similar spatial extents.
M82, with a particularly strong outflow, was shown to expel extraplanar ionized gas out to 3 kpc from its central nucleus (\citealt{Shopbell...1998...M82...ionized}, \citealt{1988...Bland...Tully...M82}).
\citealt{2011...Westmoquette...NGCoutflow..0.6} detected an ionized outflow along the minor axis of NGC 253 out to 0.6 kpc from the galactic center, similar to the lower end of spatial scales in our LDT observations.

The relative extents between \ha\ and \oiii\ for individual galaxies depends on the classification of that particular subset of SPOGs.  SF- and LINER-classified SPOGs with extraplanar emission are all detected in \ha\ and occasionally in \oiii.  Of those that show both emission lines, the \oiii\ more often fades first, extending 80-100\% as far as the \ha\ emission.  
In contrast, the AGN-classified SPOGs nearly all host \oiii\ emission that extends out to comparable distances to their \ha\ emission, and in several cases, beyond. 
Stronger \oiii~emission is expected because \oiii\ is commonly ionized by AGN (\citealt{Shopbell...1998...M82...ionized}, \citealt{woo2016}, \citealt{2020ApJ...Liu...outflows}).

In cases where the extraplanar \ha\ detections in SPOGs are coming from the eDIG around a galaxy, the detected spatial extent still matches the literature. 
Papers focused on eDIG at different scale heights show it detectable on the 1-2 kpc scale (\citealt{2000...Rossa...edig...2kpc}), out to 4 kpc above the galactic center (\citealt{2019ApJ...edig...Vel.Disp...20-60...Boettcher}), and past that to heights in the tens of kpc (\citealt{2003...Miller...edig...17kpc}).
This matches the same scale at which we see diffuse ionized gas in the area around a galaxy---from 1 kpc out to 5.2 kpc from the galactic center.

\begin{table*}
\begin{tabular}{llllllll}
\hline\hline
Galaxy name & \ha\ extent & \oiii\ extent &  Internal $\sigma$ & External $\sigma$ & Extraplanar  & Mass Loading\\
(J2000) & (kpc) &  (kpc) & (km s$^{-1}$) & (km s$^{-1}$) & Mass ($M_{\odot}$) & Factor \\
(1) &  (2) & (3) & (4) & (5) & (6) & (7) \\
\hline
\textit{AGN} & & & & & \\
J0004-0114 & 6.2 & 4.5  & 153 & 169 & 6.3 $\times$ $10^5$ & 3 $\times$ 10$^{-3}$ \\
J0102-0052 & 2.7 & 2.7  & 132 & 136 & 2 $\times$ $10^4$ & 4 $\times$ 10$^{-3}$\\
J0755+3929 & 2.9 & 2.9  & 225 & 244 & 3.7 $\times$ $10^5$ & 2 $\times$ 10$^{-3}$\\
J0800+3743 & 2.8 & 3.9  & 137 & 125 & 3 $\times$ $10^4$ & 10$^{-3}$\\ 
J0815+3720 & 2.1 & ---  & 75 & 167 & 1.7 $\times$ $10^4$ & 10$^{-3}$\\
J1007+3919 & --- & 2.1  & 232 & 187 & --- & --- \\
J1016+1323 $^{W}$ & 1.1 & 1.8 & 80 & 84 & 2.2 $\times$ $10^3$ & 10$^{-4}$\\
J1025+1647 & --- & 1.9  & 65 & 81 & --- & ---\\
J1334+3411 & 1 & 1  & 121 & 117 & 2.9 $\times$ $10^3$& 1.5 $\times$ 10$^{-3}$ \\
J1358+3901 & 2 & 2 & 84  & 144 & 3.8 $\times$ $10^4$ & 10$^{-3}$\\
J1405+1146 & 1 & 1  & 57 & 86 & 2.5 $\times$ $10^3$ & 1.7 $\times$ 10$^{-3}$\\
J1411+2531 & 1.7 & ---  & 168 & 162 & 4.8 $\times$ $10^4$ & 1 $\times$ 10$^{-3}$\\

\hline
\textit{SF} & & & & & \\
J0141-0015 $^{\dag}$ & 2.2 & --- &  40 & 26 & $^{\dag}$  & $^{\dag}$\\
J0204+0051 $^{W}$ & 5.9 & ---  & 64 & 78 & 1 $\times$ $10^5$ & 1.2 $\times$ 10$^{-5}$\\
J0728+3654 $^{\dag}$& 2.4 & ---  & 93 & 67 & $^{\dag}$  & $^{\dag}$\\
J0813+2434 $^{W}$ & 1.1 & --- & 92 & 103 & 2 $\times$ $10^3$ & 2.4 $\times$ 10$^{-6}$\\
J0826+4558 & 0.7 & ---  & 184 & 226 & 5 $\times$ $10^5$ & 4.5 $\times$ 10$^{-3}$\\
J0843+2205 $^{\dag}$& 1.8 & 1.7  & 61 & 37 & $^{\dag}$ & $^{\dag}$\\
J0937+3335 $^{\dag}$ & 1.8 & ---  & 54 & 50 & $^{\dag}$ &  $^{\dag}$\\
J1148+5459 $^{\dag}$ & 0.9 & 0.7  & 69 & 59 & $^{\dag}$  & $^{\dag}$\\
J1307+5350 $^{W}$ & 2.5 & ---  & 81 & 90 & 1.7 $\times$ $10^4$& 6.6 $\times$ 10$^{-5}$ \\
J1439+5303 & 2 & ---  & 70 & 155 & 5.8 $\times$ $10^3$ & 3.4 $\times$ 10$^{-5}$\\
J1540+5106 $^{\dag}$ & 2 & ---  & 47 & 44 & $^{\dag}$  & $^{\dag}$\\
J1702+3254 $^{\dag}$ & 5.2 & ---  & 120 & 97 & $^{\dag}$  & $^{\dag}$ & \\
J1703+2531 $^{\dag}$ & 2.8 & 2.4 & 53 & 50 & $^{\dag}$ & $^{\dag}$ \\
J2129-0010 & 2.5 & ---  & 69 & 100 & 1.2 $\times$ $10^4$ & 4.2 $\times$ 10$^{-5}$\\
J2300+0036 & 3.1 & 3.1  & 60 & 83 & 7.4 $\times$ $10^4$&  1.2 $\times$ 10$^{-4}$ \\

\hline
\textit{LINER} & & & & & \\
J0142+1309 & 1.3 & ---  & 214 & 252 & 8.1 $\times$ $10^2$ & 4 $\times$ 10$^{-4}$\\
J0821+2238 $^{\dag}$ & 2.7 & 2.2 & 53 & 58 & $^{\dag}$ & $^{\dag}$\\
J1007+3240 & 2.4 & ---  & 157 & 171 & 6.1 $\times$ $10^3$ & 8.2 $\times$ 10$^{-5}$\\
J1156+2829 & 2 & --- &  62 & 130 & 2 $\times$ $10^3$ &  0.3\\

\hline\hline
\end{tabular}
\caption{Table showcasing the furthest detectable extent of \ha\ and \oiii\ extraplanar emission, velocity dispersions internal and external to the galaxy, and estimated gas masses based on \ha\ and \oiii.
$^{\dag}$ marks galaxies with velocity dispersion consistent with eDIG.
$^{W}$ marks galaxies with velocity dispersion consistent with a weak outflow.  
Mass-loading factors are described in Section~\ref{massloading}.
} 
\label{mass_galaxy_property_table}
\end{table*}

\subsection{What is Ionizing the Extraplanar Gas?}
\label{what is ionizing}

\begin{figure*}[ht]
\centering
  \subfigure{\includegraphics[width=0.45\textwidth, trim=0cm 0cm 0cm 0cm,clip]{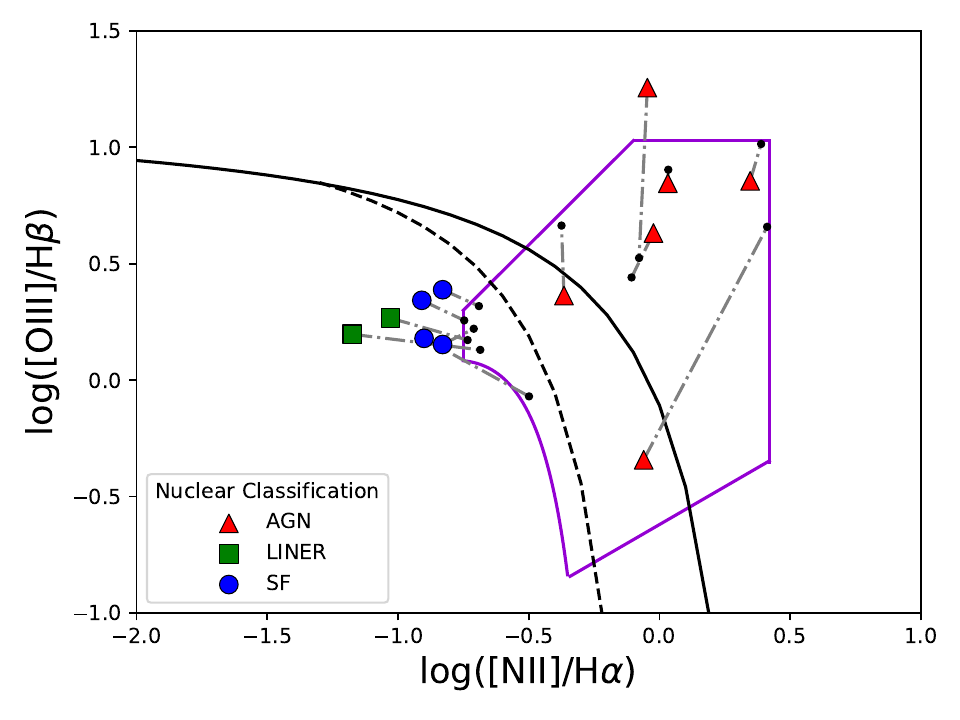}}
  \subfigure{\includegraphics[width=0.45\textwidth, trim=0cm 0cm 0cm 0cm,clip]{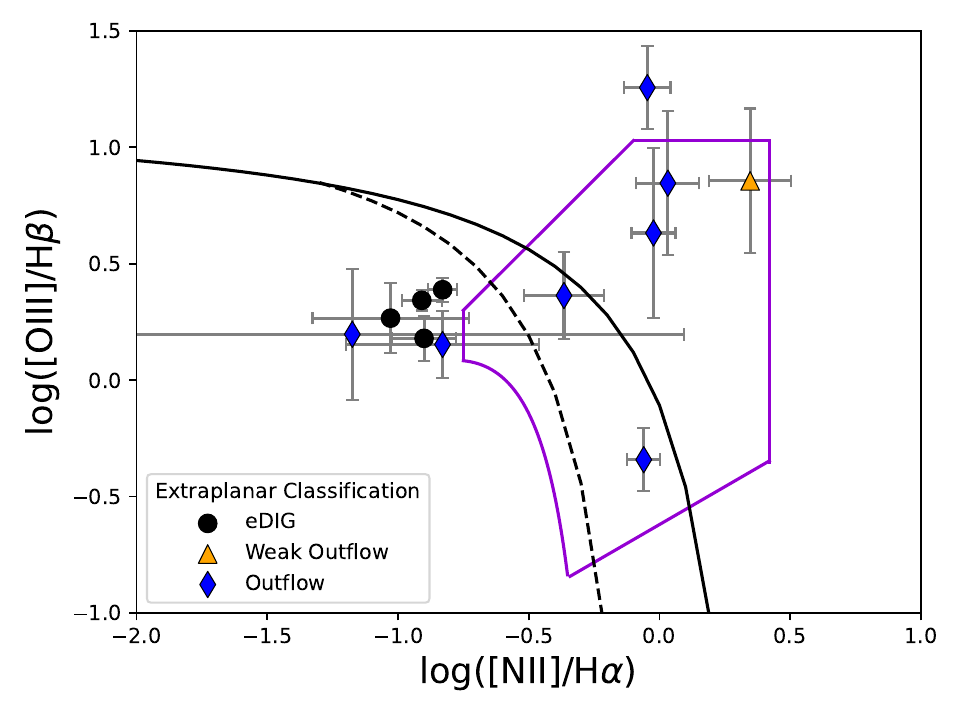}}
\caption{BPT line diagnostic diagrams of the extraplanar ionized gas detected in our SPOGs sample.
For each galaxy, the extraplanar gas is summed together for each emission line and then plotted as a single data point. 
Overlaid are the star formation boundaries from \citealt{2006...Kewley...BPT...SFcurve} (black line) and \citealt{Kauffmann...2003} (dashed line).
\textbf{Left:} Extraplanar gas data points are different shapes/colors based on the central galactic classification based on nuclear fiber data taken from OSSY  (\citealt{Oh...2011}): red triangles denote an AGN nuclear classification (blue circles denote SF and green squares denote LINER). 
A gray dot and dashed line connects the extraplanar gas points to their nuclear fiber data (black dot).
\textbf{Right:} Extraplanar gas data points are different shapes/colors based on their kinematic classification. 
Black circles mark low velocity dispersion extraplanar as consistent with eDIG.
Yellow triangles (Weak outflow) and blue diamonds (outflows) mark extraplanar gas with higher velocity dispersions than the gas internal to the galaxy.
Gray error bars are plotted for the extraplanar gas data.
AGN extraplanar ionized gas is consistent with AGN as their main power source. 
SF and shock-heated ionized gas have lowered line ratios than the shock model boundaries used in part to select SPOGs.
} 
\label{spogsbpt}
\end{figure*}

Figure~\ref{spogsbpt} uses the BPT diagnostic diagram to look at the detected extraplanar gas in our LDT SPOGs sample. 
The left panel plots summed line ratios for each galaxy as a single data point.
Each point is colored to match the galaxy's nuclear fiber classification (\textsection \ref{classification}).
Red triangles, blue circles, and green squares denote AGN, star-forming, and shock-heating (LINER) dominated SPOGs, respectively.
For 19 out of \extraplanar{} SPOGs with confirmed extraplanar gas, the extraplanar ionized gas is only detected in one or two emission lines, and those galaxies are excluded from this plot. 
However, in the strongest 12 cases, we detect \hb\ and \nii\ and investigate the ionization diagnostics using the BPT diagnostic diagram (\citealt{1981...BPT...NII}).

The right panel of Figure~\ref{spogsbpt} plots the same SPOGs galaxies with extraplanar gas. 
In this case, points are marked by whether or not the extraplanar gas is classified as eDIG, weak outflow, or outflowing.
Most outflowing extraplanar gas lies within the composite and AGN region of the diagnostic plot. 

Extraplanar ionized gas generally matches the respective SPOGs' nuclear classification.
Moreover, extraplanar gas in 7 SPOGs falls outside of the SPOGS shock boundary from \citealt{Alatalo..SPOGs..2016A}.
As expected of the AGN SPOGs population, extraplanar gas consistently have elevated line ratios in the BPT diagram above the \citealt{Kauffmann...2003} star formation boundary.
It is likely that the central AGN is the main power source for this outflowing gas. 
Of interest are the SF and LINER galaxy points.
The 2 LINER SPOGs have extraplanar gas that lies closer to the SF sequence on the BPT.
All extraplanar emission classified as eDIG in Section~\ref{kinematics and outflows? eDIG?} shows ionization consistent with \ion{H}{2} regions, as expected.
However, two galaxies have extraplanar gas consistent with our outflow definition but ionization ratios along the SF sequence of the BPT.
\citealt{Alatalo..SPOGs..2016A} mentioned that some subsample of SPOGS at low redshift might be interloper blue spiral galaxies that are not actually shocked post-starbursts. 
Future work using the long-slit spectra along the major axis will better diagnose whether or not the SF SPOGs shown in Figure \ref{spogsbpt} truly are SPOGs.

\section{Mass Entrained in Extraplanar Gas}
\label{mass_section}

For galaxies that have outflowing gas, \ha\ traces the entrained gas mass being pushed out of the galaxy by an outflow.
We convert \ha\ extraplanar flux to luminosity using the luminosity distance to each galaxy.
We then use the cosmology calculator of \citealt{cosmology_wright} along with redshifts obtained from SDSS (\citealt{SDSS..DR7}) to calculate the luminosity distance.
Due to the low signal of extraplanar \hb, we are not correcting for dust attenuation using the Balmer decrement.
However, little dust is expected to be in the extraplanar ionized gas as compared to the galactic plane, so this should not affect our results.

We determine mass in the outflows using an equation following \citealt{2006agna.book.....Osterbrock} and \citealt{2020ApJ...Liu...outflows}: 

\begin{equation}
M_{\text{out}} = 4.48  M_{\odot} \times \left( \frac{L_{\text{Ha}}}{10^{-35} \, \text{erg} \, \text{s}^{-1}} \right) \left( \frac{\langle n_e \rangle}{100 \, \text{cm}^{-3}} \right)^{-1}
\end{equation}

where $\langle n_e \rangle$ is the electron density and $L_{\text{Ha}}$ is the \ha\ luminosity calculated and summed over the entire extraplanar gas extent. 

The ratio of the \sii\ doublet would give a direct measurement of the electron density.
Unfortunately, for our sample, \sii\ is not sufficiently bright in our extraplanar gas for each line individually to meet a signal-to-noise threshold of 3.
Instead, we calculate our masses using electron densities that have been measured in other extraplanar ionized gas studies.

M82 is a prototypical starburst galaxy showing strong outflows that has been well-studied; its outflow has an electron density ranging from 200 to 400 cm$^{-3}$ (\citealt{2009...Westmoquette...M82} and \citealt{2011...Yoshida....M82}).
\citealt{2014...Ho...SAMI...electron} found a similar electron density in outflowing gas of a normal star-forming galaxy.
Thus, we adopt 300 cm$^{-3}$ as our electron density for outflowing extraplanar gas.
In the case of weak outflows, we use the same electron density as for the outflowing gas.

We calculate the external gas masses of our detected outflows, given in Table~\ref{mass_galaxy_property_table}. 
The masses in this table are the total extraplanar gas mass estimated using the sum of gas mass per pixel.
We do not estimate the mass of extraplanar ionized gas classified as eDIG because the gas is not being removed from the system, and therefore unlikely to be relevant to the quenching processes of our galaxies.  
The electron density for ionized gas considered eDIG is considerably lower than the electron density of outflowing gas.
Measurements of warm ionized gas extending out past the galactic limits result in electron densities of 0.03-0.22 cm$^{-3}$ (\citealt{1973...Reynolds....electrondensity...edig}, \citealt{1991...edig...electron...density...0.08}, \citealt{2008...edig....elecdens....0.03}).

From Table \ref{mass_galaxy_property_table}, it is clear that LINER SPOGs represent a lower end of entrained ionized gas found in the outflows on the order of $10^3$ $M_{\odot}$. 
In comparison, AGN and SF SPOGs host outflows with larger gas masses on the order of $10^3$ to $10^5$ $M_{\odot}$.

\citealt{2020ApJ...Liu...outflows} found ionized gas masses for the outflowing component in the range of 10$^5$ to 10$^{7.3}$ M$_\odot$ for dwarf galaxies with AGNs.
In cases of edge-on starburst galaxies, ionized gas masses on the order of 10$^8$ M$_\odot$ are detected (\citealt{1996...Lehnert...Starburst....Heckman...edge-on}). 
On the other hand, the outflows presented here show ionized gas masses significantly lower -- but are themselves lower limits.  The true gas masses entrained may be higher if the outflow extends outside our slit width, or if the gas is diffuse enough to remain below our surface brightness detection limit.  If our measurements are representative, then these outflows are only a small fraction $<$1\% of the total gas contents of the galaxies and may not be significant for quenching. 
There is growing evidence that poststarburst galaxies retain a significant amount of gas mass \citep{2016...Alatalo...2016b...Co...gas, 2022...Smercina...PSBs...gas}, so it is reasonable that mass entrained in the outflows might be small.

Additionally, the outflow mass traced by ionized gas is likely a lower limit of the true multi-phase outflow gas content. 
Neutral and molecular gas phases likely are much more massive than the content of the ionized extraplanar gas.
Work done by \citealt{2013...Rupke...Outflows...v98...gasphases}, \citealt{2015...Carniani....gasphase...ionizedvsmolecular}, and \citealt{2021...Fluetsch...outflow...outflowrate....gasphases} found that the contributions from neutral and/or molecular gas accounted for 60-90\%\ of outflow rates in the galaxies they studied. 
When looking at post-starburst E+A galaxies, the same trend appears to hold. 
For a sample of 144 post-starbursts, \citealt{2022...PSBs...gas...neutral....gasphases...Baron} found that the neutral gas---when detected---held 10-100 times more mass than the ionized gas phase.

\section{Discussion}

With a division between outflows and eDIG extraplanar gas, our next task is to quantify the effectiveness of the SPOGs outflows at removing gas from their hosts. 
To do so, we are interested in finding a mass loading factor ratio of outflow rate, $\dot{M}_{out}$, to accretion rate, $\dot{M}_{acc}$ (for AGN) or to star formation rate  $\dot{M}_{SFR}$ (for SF and LINER galaxies), to note the efficiency of the outflow in driving gas out of any SPOG with outflowing extraplanar gas.

\subsection{Outflow Timescales} 
Outflow rates can be estimated from the extraplanar ionized gas mass and timescale over which that outflow occurred. 
Unfortunately, most observed galaxies show no significant velocity offset between the extraplanar gas and the host galaxy from which to estimate a timescale.
This is most plausibly due to the inclination angle of most galaxies and our sample targeting edge-on SPOGs (see Figure \ref{2d_spectral_emission}).
Even strong outflows, if in the plane of the sky, would show no radial velocity offsets from the systemic velocity of the galaxy.

There do exist galaxies in our sample that can provide kinematic information, given that the system is not perfectly edge-on. 
J0800+3743 is such a galaxy.
It is a slightly inclined AGN SPOG with an inclination angle of 82\arcdeg.
The furthest point in the ionized extraplanar gas has a velocity of 210 km s$^{-1}$. 
When looking for a timescale for this outflow, the furthest point gives us an upper limit on the detected outflow if the ionized gas can be assumed to slow down.
For gas to reach out to 3.9 kpc (see table \ref{mass_galaxy_property_table}) with a velocity of 210 km s$^{-1}$, it would take 1.8 $\times 10^7$ years.  

Alternatively, we consider the post-burst age for each SPOG as a timescale from which we can derive outflow rates. Crucially, we use this for both AGN and SF SPOGs, operating on the assumption that AGN activity was likely fueled by the same gas that fueled the starburst.
Although \citealt{2007...Davies...SFAGN...timescale} showed a delay of 50-100 Myr between AGN and SF timescales in their sample of Seyfert galaxies, our $t_{burst}$ is generally much longer (200-1000 Myr), so our assumption that $t_{AGN}$ $\sim$ $t_{burst}$ is likely reasonable.
We fit the $t_{burst}$ using a method adapted from \citealt{2018...Decker...spogs...postburstage} and using Bagpipes (\citealt{2018...Carnall...2018....Bagpipes}, \citealt{2019...Carnall}). 
The star formation histories are modeled as a young exponential burst population, with age, burst duration, and mass varying with a log10 prior. 
The old stellar population is modeled as a delayed exponential with duration 1 Gyr and variable age. 
This method makes three substantive changes to the method in \citealt{2018...Decker...spogs...postburstage}. 
The first is the variation of the age of the old stellar population. 
The second is that the metallicity is varied using a Gaussian prior, with center and width from \citealt{2005...Galazzi....SDSS...metallicities} and stellar masses from MPA-JHU (\citealt{Kauffmann...2003}, \citealt{2004...Brinchmann...2004...SDSS...study}, \citealt{Tremonti....2004....SDSS...MPAJHU}). 
The third is that the entire spectrum, rather than Lick indices and photometry, is fit, enabled by the use of a Gaussian process noise model as implemented in \citealt{2019...Carnall}. 
The dust attenuation is varied, assuming a Calzetti dust law (\citealt{2000...Calzetti}). 
Other varied parameters are the redshift, velocity dispersion, and noise scaling. 
We assume a Kroupa IMF (\citealt{2002...Kroupa...IMF}).
 
To obtain outflow rates, we use the extraplanar gas mass calculated in Section~\ref{mass_section} and the calculated time since the starburst began as a consistent timescale.
We note that this time estimate provides a lower limit on outflow rate, because some or all outflows may not be launched until after the starburst began.
For comparison, our galaxy with a measured outflow velocity J0800+3743 has a measured time since starburst of 2.2 $\times 10^8$ years, a factor of about ten higher than our outflow velocity and distance calculation.  
If the outflow timing relative to starburst of J0800+3743 is typical, then our calculated mass-loading factors can be taken as lower limits.

\subsection{Calculating Mass-Loading Factors}
\label{massloading}

To estimate mass-loading factors for AGN outflows ($\eta_{AGN} \equiv \dot{M}_{out}/\dot{M}_{acc}$), we use the accretion rate estimate from \citealt{2002...Yu...Tremaine...Mdot} (and references therein):

\begin{equation}
    \dot{M}_{acc} = \frac{L_{bol}}{\epsilon * c^2}
\end{equation}

where $L_{bol}$ is the bolometric luminosity, $\epsilon$ is the radiative efficiency, and c is the speed of light. We estimate the bolometric luminosity from \oiii~luminosity with the correction factor of 3500 $L_{\oiii} = L_{bol}$ used by \citealt{2004...Heckman...Kauffmann....3500bol...oiii}.  Here we use \oiii~luminosity from the OSSY catalog \citep{SDSS..DR7, Oh...2011}, corrected for extinction using the Balmer decrement.

To estimate mass-loading factors for SF and LINER galaxies hosting outflows, we compare our outflow rate to star formation rates instead: ($\eta_{SF} \equiv \dot{M}_{out}/SFR$). 
As a comparison, we use two different star formation rates in this calculation: the current SFR (SFR$_{H\alpha}$) based on H$\alpha$ luminosity and the averaged SFR (SFR$_{averaged}$) since the last starburst.
In the first case, we use $L_{H\alpha}$ from OSSY (\citealt{Oh...2011}) to probe current star formation.
We convert $L_{H\alpha}$ to SFR using the conversion factor 7.9 $\times$ 10$^{-42}$ M$_{\odot}$ yr$^{-1}$ (ergs s$^{-1})^{-1}$ (\citealt{1994...Kennicutt...SFR}). 
We then convert to a \citealt{2003..Chabrier...IMF} initial mass function by dividing by 1.53.
In the second case, instead calculate the average SFR since the burst.
We fit star-formation histories using the method mentioned earlier adapted from \citealt{2018...Decker...spogs...postburstage}.  
We estimate the averaged SFR, SFR$_{averaged}$, as the mass formed in the burst divided by the time since the burst began.  

We report our mass-loading factors for all galaxies in Table~\ref{mass_galaxy_property_table} and discuss them for each class of galaxy below.  

\subsection{AGN-host SPOGs}

A higher fraction of AGN-host SPOGs (52.2\%) show detected extraplanar ionized gas than either LINER or SF SPOGs (20\% and 44.1\%, respectively).  As might be expected, 2/3 of the detected extraplanar \oiii\ emission across our sample is found in AGN hosts.  All AGN hosts with strong enough outflows to classify show composite- or AGN-like line ratios in the extraplanar gas too, suggesting that the dominant ionization source in the outflow is photoionization from the AGN.

All of the ionized gas in AGN SPOGs is classified as either weakly outflowing or as an outflow using the velocity dispersion of the extraplanar gas, higher than any other classification (see Table \ref{outflow_fraction_table}).
AGN host galaxies also carry more mass in their ionized gas outflows than either of the other classifications, as listed in Table \ref{mass_galaxy_property_table}, to the furthest spatial extents.  

Mass-loading factors for our AGN-driven outflows typically are on the order of $10^{-3}$, listed in Table \ref{mass_galaxy_property_table}.  
These low values (much below 1) suggest that our outflows are not effective at driving gas out of the system, even if our ionized gas detections only trace $\sim$ 1\%\ of the full multiphase outflow.

\begin{figure*}
\centering
\includegraphics[scale=0.55, trim=0cm 0cm 0cm 0cm,clip]{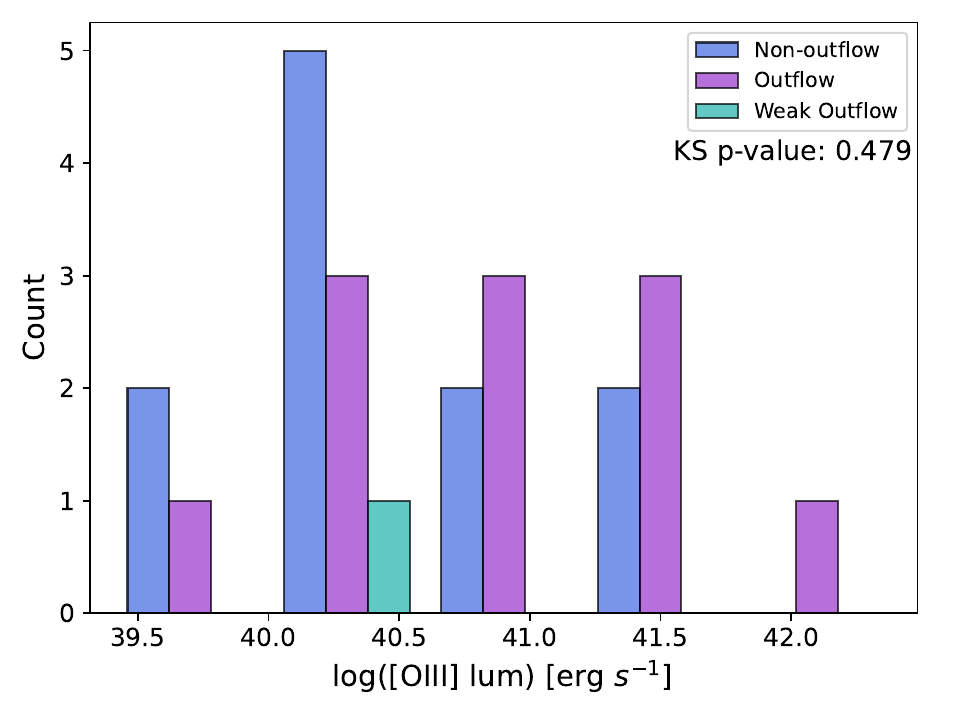} \\
\caption{Histogram of the AGN SPOGs population observed divided in bins of nuclear \oiii\ luminosity, extinction-corrected from the OSSY catalog \citep{SDSS..DR7, Oh...2011}.
The population was split into AGN SPOGs with confirmed \oiii\ outflows (purple), weak outflows (green), and those that did not have confirmed outflows (blue).  A Kolmogorov-Smirnov test shows that the two samples are not likely drawn from different distributions.
AGN power is not the dominant factor in determining whether outflows are launched in this sample.
}
\label{agn_hist}
\end{figure*}

Figure \ref{agn_hist} shows the AGN SPOGs population separated into outflowing (purple), weakly-outflowing (green), and non-outflowing (blue) distributions in bins of central \oiii\ luminosity, extinction-corrected from the OSSY catalog \citep{SDSS..DR7, Oh...2011}. 
The outflowing sample skews a little to lower luminosities than the non-outflow population, but overall the two populations are not significantly distinct. 
A Kolmogorov-Smirnov test gives a $p$-value of 0.479, suggesting that \oiii\ luminosity (i.e. AGN accretion rate) is not a good indicator of outflow likelihood.  

\begin{figure}
 \gridline{\fig{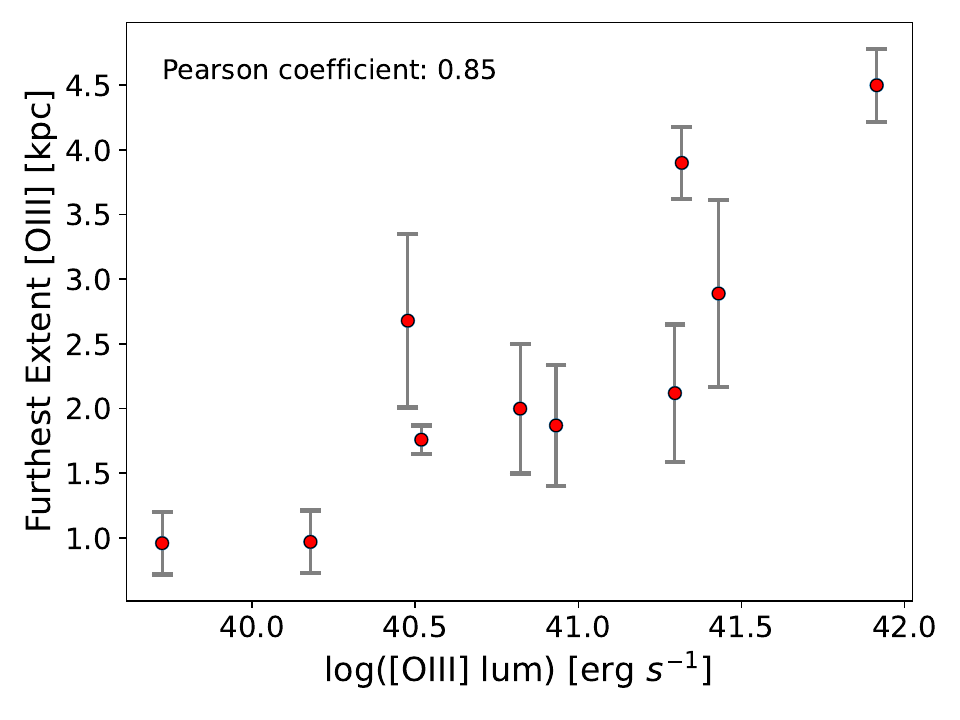}{0.45\textwidth}{}
  \hspace{-2.2cm} 
  \fig{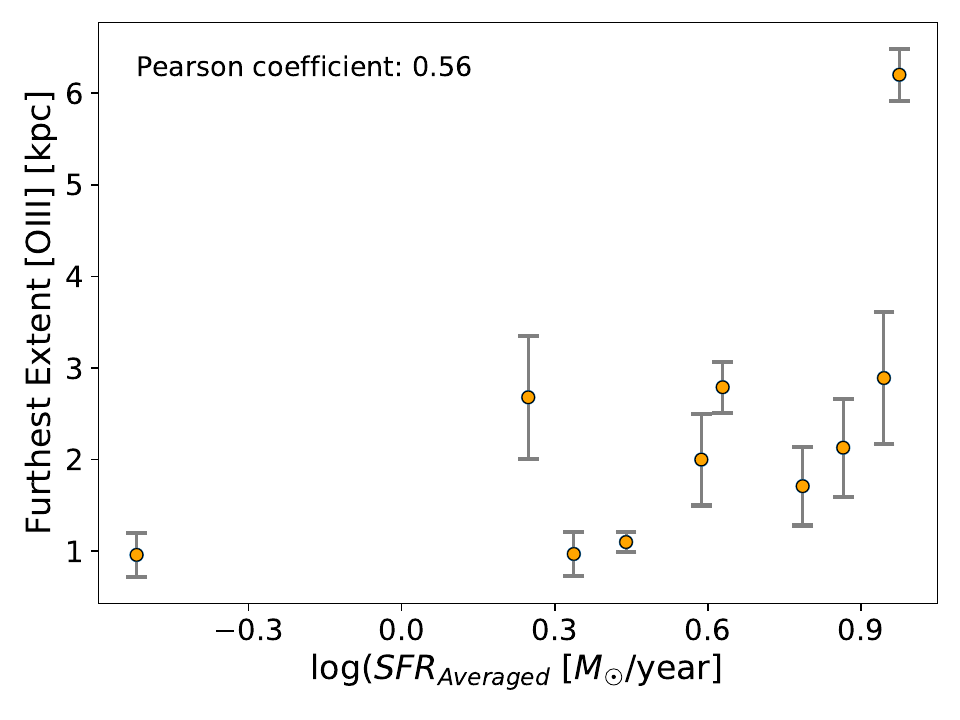}{0.45\textwidth}{}}
  \vspace{-1cm} 
  \gridline{\fig{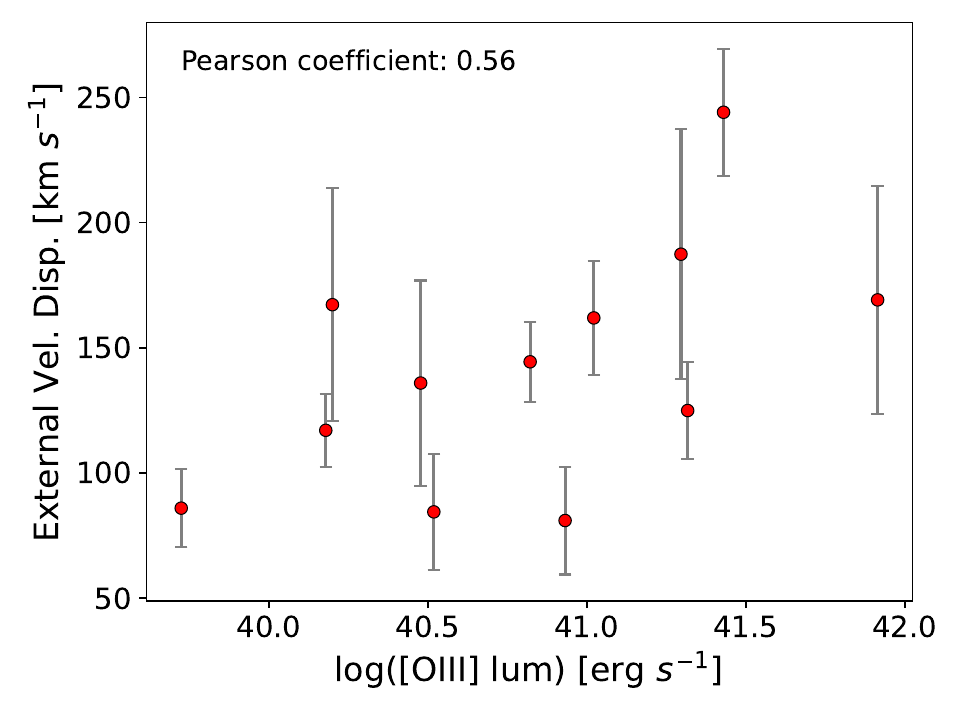}{0.45\textwidth}{}
  \hspace{-2.2cm} 
  \fig{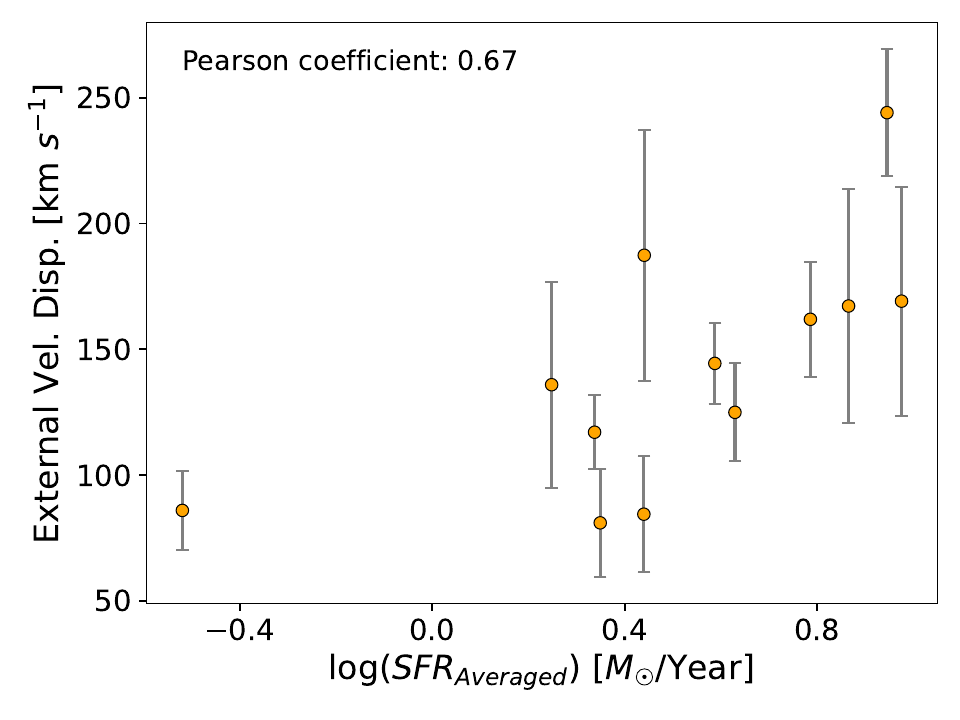}{0.45\textwidth}{}}
  \vspace{-1cm} 
  \gridline{\fig{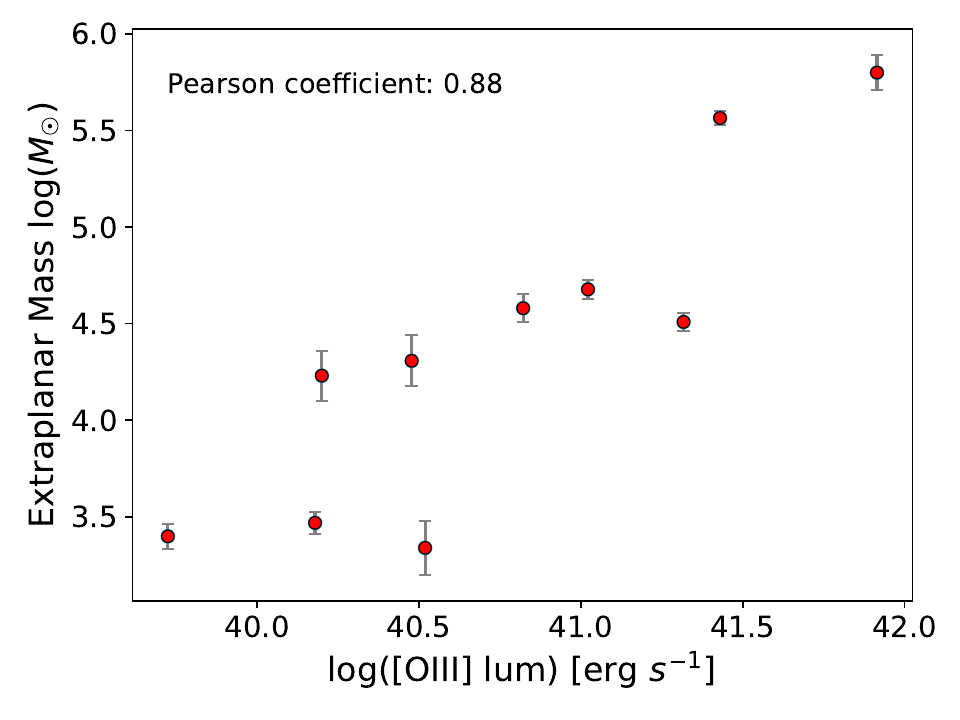}{0.45\textwidth}{}
  \hspace{-2.2cm} 
  \fig{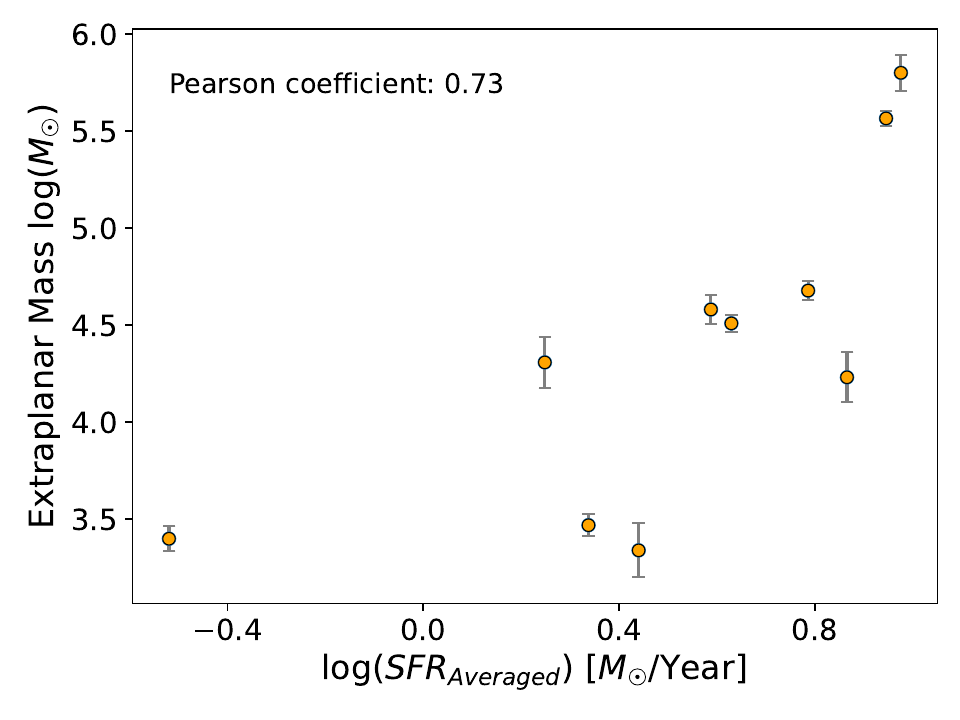}{0.45\textwidth}{}} 
  \caption{Outflow characteristics for AGN SPOGs as a function of relevant driving power: log(L$_{\oiii}$) as a proxy for AGN luminosity (red points, left column) and log(SFR$_{averaged}$) for recent starburst activity (orange points, right column).  
  The top row shows the furthest detectable extent of ionized gas.
  The second row shows the average external velocity dispersion of the extraplanar gas.
  The third row shows total extraplanar ionized gas mass.
  Each panel also includes the Pearson correlation coefficient.  
  The furthest extent is correlated closer with AGN luminosity.
  Velocity dispersion shows a slightly stronger correlation with SFR$_{averaged}$.  
  Extraplanar gas mass is positively correlated with AGN power, more so than SFR$_{averaged}$.  
  From all three outflow gas parameters, it looks likely that in AGN SPOGs, the central AGN has driven these outflows.
  }
  \label{fig:six-panel-lum analysis}
\end{figure}

In the AGN SPOGs, two of the most likely outflow drivers are the central AGN and the recent starburst. 
We examine their properties (furthest extent, external velocity dispersion, and extraplanar mass) as a function of central \oiii\ luminosity and $SFR_{Averaged}$ as proxies for AGN accretion rate (left column of Figure \ref{fig:six-panel-lum analysis}) and recent starburst activity (right column of Figure \ref{fig:six-panel-lum analysis}), respectively.  
Pearson correlation coefficients are given in the top left of each panel. 
In the cases of furthest extent and extraplanar gas mass, there are noticeable linear trends suggesting more luminous AGN power outflows that are faster and carry more mass out of said galaxies. 
This is in contrast with a weaker correlation between the extent of the extraplanar gas and $SFR_{Averaged}$. 
For external velocity dispersion, there is no strong correlation with either \oiii\ luminosity or $SFR_{Averaged}$, only a positive one that is not significantly different between either ionization source.
Although the outflow properties do correlate with $SFR_{Averaged}$, the correlations with \oiii\ luminosity are marginally stronger, suggesting that the central AGN is more likely the driving source of outflows in the AGN SPOGs.
These trends are in line with previous studies \citep[e.g.][]{2017...concas...OIII}, showing that AGN feedback is a significant mechanism in driving ionized outflows.

\begin{figure*}
  \centering
  \subfigure{\includegraphics[width=0.5\textwidth, trim=0cm 0cm 0cm 0cm,clip]{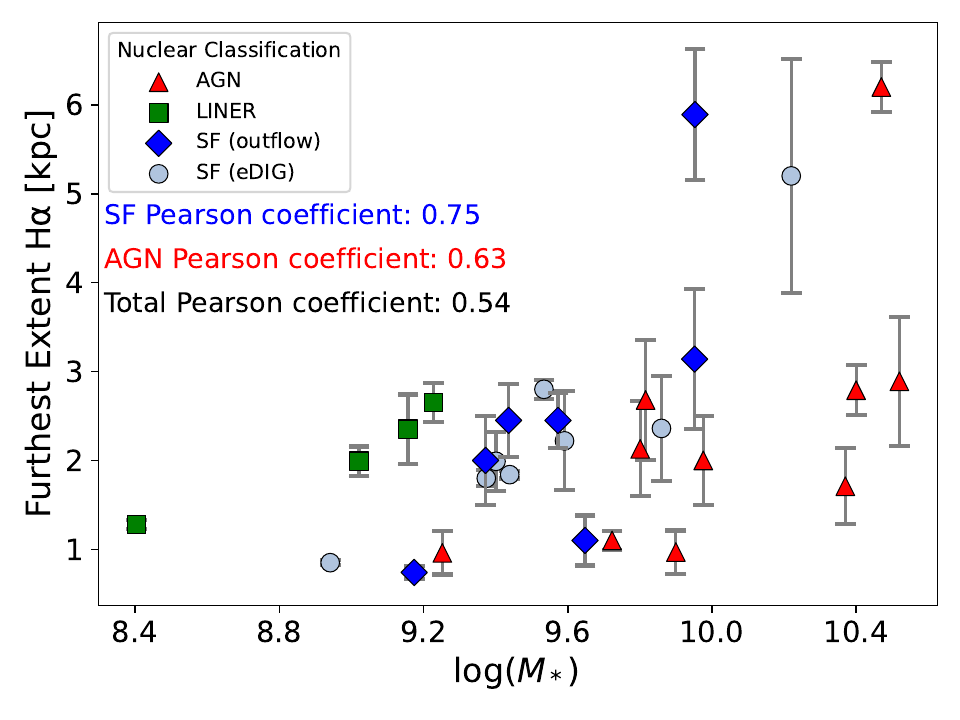}}
  \subfigure{\includegraphics[width=0.5\textwidth, trim=0cm 0cm 0cm 0cm,clip]{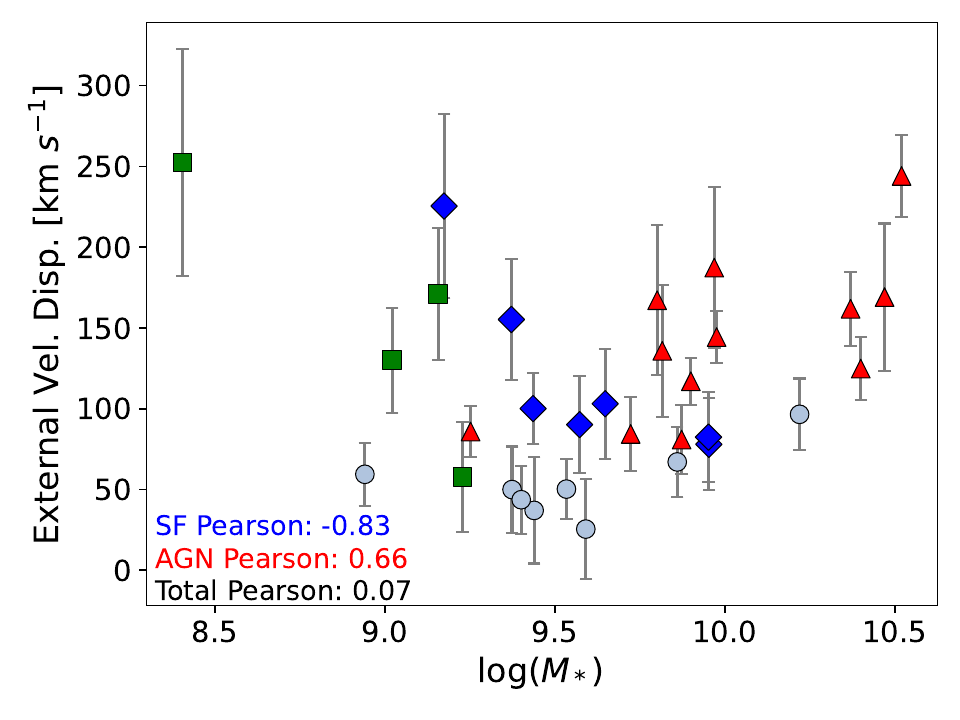}}
  \subfigure{\includegraphics[width=0.5\textwidth, trim=0cm 0cm 0cm 0cm,clip]{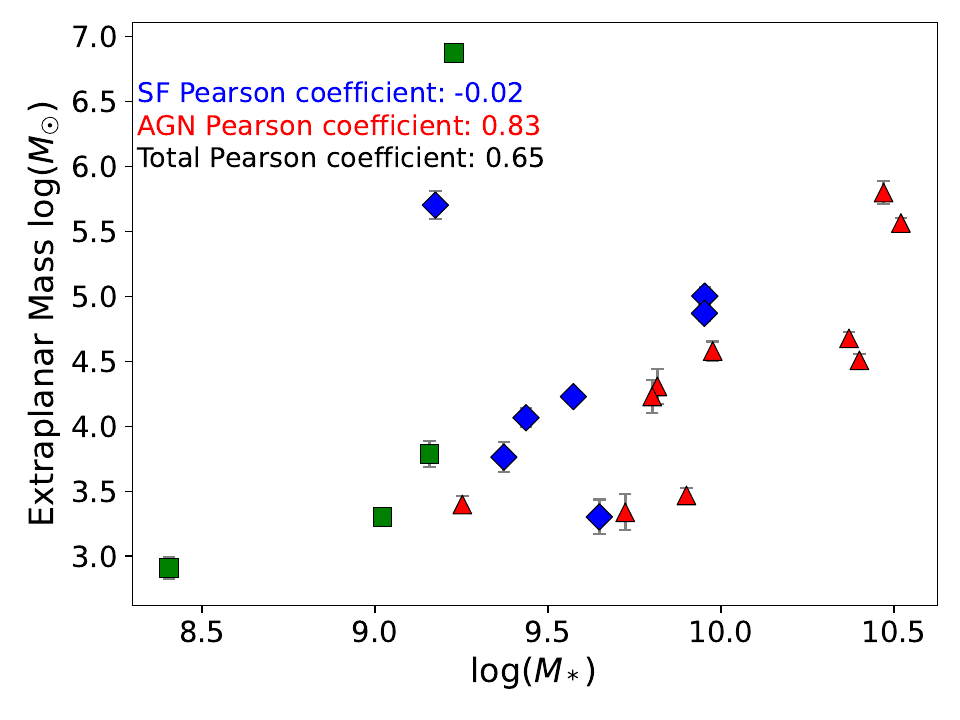}}
  \caption{Outflow characteristics as a function of host galaxy stellar mass: furthest detectable extent of ionized gas (top panel), average external velocity dispersion of the extraplanar gas (middle panel), and total extraplanar ionized gas mass (bottom panel).
  Each panel also includes the Pearson correlation coefficient for each galaxy classification.  (Note that eDIG points are plotted in the SF panels for reference, but \textit{not} included in the Pearson correlation tests.)
  Outflow properties are correlated with galaxy stellar mass for AGN hosts, albeit slightly less strongly than with AGN power.  
  For starburst-driven outflows, stellar mass correlates with furthest extent, anticorrelates with external velocity dispersion, and does not correlate with extraplanar mass.
  }
\label{Mass Characteristics}
\end{figure*}

We also examine the same outflow characteristics as a function of galaxy stellar mass, shown for AGN as red triangles in Figure~\ref{Mass Characteristics}.  Furthest extent, external gas velocity dispersion, and extraplanar gas mass all correlate moderately well with AGN power.  
AGN luminosity tends to correlate with host galaxy stellar mass (i.e. bigger galaxies can drive stronger AGN), so it's not surprising that the correlation persists.  
However, we note that the Pearson correlation coefficients are higher (so the correlation is stronger) between outflow characteristics and L$_{\oiii}$ than with stellar mass.

\subsection{LINER host SPOGs}
In contrast, LINER SPOGs have fewer extraplanar detections in our observing sample: only 20\% of these galaxies host extraplanar ionized gas. 
They make up the lowest outflow fraction of detected ionized gas for \ha\ and \oiii, and 1/4 of these are classified as eDIG based on gas kinematics. 
Additionally, although their external velocity dispersion matches those of the AGN outflows, they do not carry as much material entrained in those outflows: an order of magnitude smaller than AGN or most SF galaxies.  
The mass-loading factors for LINER host SPOGs are all below one (and half below $10^{-4}$), indicating that the starburst-driven outflow is not likely a dominant factor in quenching. 

Galaxies classified as LINERs are now understood to be a mixed bag of different physical mechanisms.  LINER-like emission can be produced by shock-ionization and therefore may be associated with galactic winds \citep[e.g.][]{2010ApJ...Rich...outflows...vel.disp,2011...Vel.Disp....Rich...elevated.line.ratios}, or low-luminosity AGN mixed with other ionization mechanisms \citep[e.g.][]{2017...Marquez...LINERS...LLAGN}.  An analysis of MaNGA data also showed that a substantial fraction of LINER-classified galaxies instead are likely dominated by hot, evolved stars \citep{2016...Belfiore...LIERs...MaNGA}.  The blend of mechanisms associated with LINER-like emission likely contributes to the low detection of extraplanar emission.  Whichever mechanisms dominate in our sample of LINERs, the outflowing gas we do detect shows emission line ratios consistent with star formation and turbulent (elevated velocity dispersion) gas kinematics.

\subsection{SF host SPOGs}

The fraction of SF galaxies in which we detect extraplanar emission is 44.1\%, nearly as high as for AGN; but, we classify 8/15 of those detections as eDIG and 3/15 of them as weak outflows.  
This outflow frequency (7/34, or 20.6\%, of SF SPOGs when including weak outflows) is significantly lower than that found for edge-on star-forming galaxies by \citealt{Ho-2016-edgeon-outflows}, who found an outflow fraction of 15/40 (37.5\%).  
It is a similar outflow frequency found by \citealt{2020...Roberts-Borsani...Manga...SF...outflow...population} for a star-forming population of galaxies when examining \ion{Na}{1} D. 
The extraplanar velocity dispersion in outflowing SF SPOGs has a median value of 100 km s$^{-1}$, which is smaller than the median velocity dispersions of 140 and 171 km s$^{-1}$ for the AGN SPOGs and LINER SPOGs outflowing populations, respectively. 
The most extreme objects, like J0826+4558, drive similar external velocity dispersions to the strongest AGN and LINER SPOGs.

\begin{figure*}
\centering
\includegraphics[scale=0.55, trim=0cm 0cm 0cm 0cm,clip]{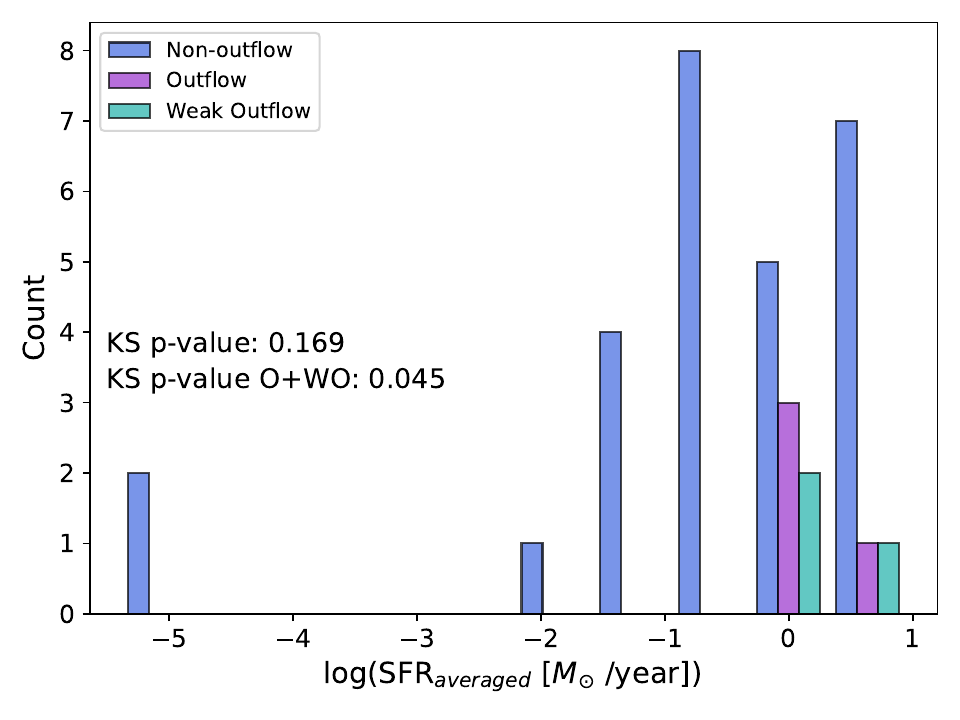} \\
\caption{Distributions of SFR for the SF SPOGs samples that host outflows (purple), weak outflows (green), and show no signs of outflow (blue).  Kolmogorov-Smirnov tests show there is no significant difference between the outflow and non-outflow samples.  We repeat the statistical test comparing the (outflow + weak outflow) sample to the non-outflow sample and find a p-value of 0.045.  
Outflows are rare in our SF SPOGs sample, and there is suggestive but not conclusive evidence that SFR$_{averaged}$ could contribute to the likelihood of driving them.
}
\label{sf_hist}
\end{figure*}

We hypothesize that the detected outflows are driven by star formation during the past starburst.  
Figure \ref{sf_hist} examines the distributions of average SFR since the burst SFR$_{averaged}$, estimated in \textsection \ref{massloading}, for both the outflowing and non-outflowing SF SPOGs populations.  
The non-outflowing sample includes both galaxies for which no extraplanar gas was detected and the galaxies hosting eDIG.
(Of note, two galaxies have significantly lower SFR compared to the rest of the sample. 
These were two SPOGs galaxies in which a recent starburst was not preferred in our star formation history modelling.)
The outflow and non-outflow host SF SPOGs do not show significantly different SFRs, with a Kolmogorov-Smirnov (KS) test showing the two populations are not likely drawn from different populations (a p-value of .17).
However, a similar KS test was done combining the outflows with weak outflows, yielding a lower p-value of 0.045.  
The decrease in p-value with increased sample size is suggestive but not conclusive evidence that a higher SFR$_{averaged}$ could contribute to the likelihood of driving outflows.

\begin{figure*}
  \centering
  \subfigure{\includegraphics[width=0.5\textwidth, trim=0cm 0cm 0cm 0cm,clip]{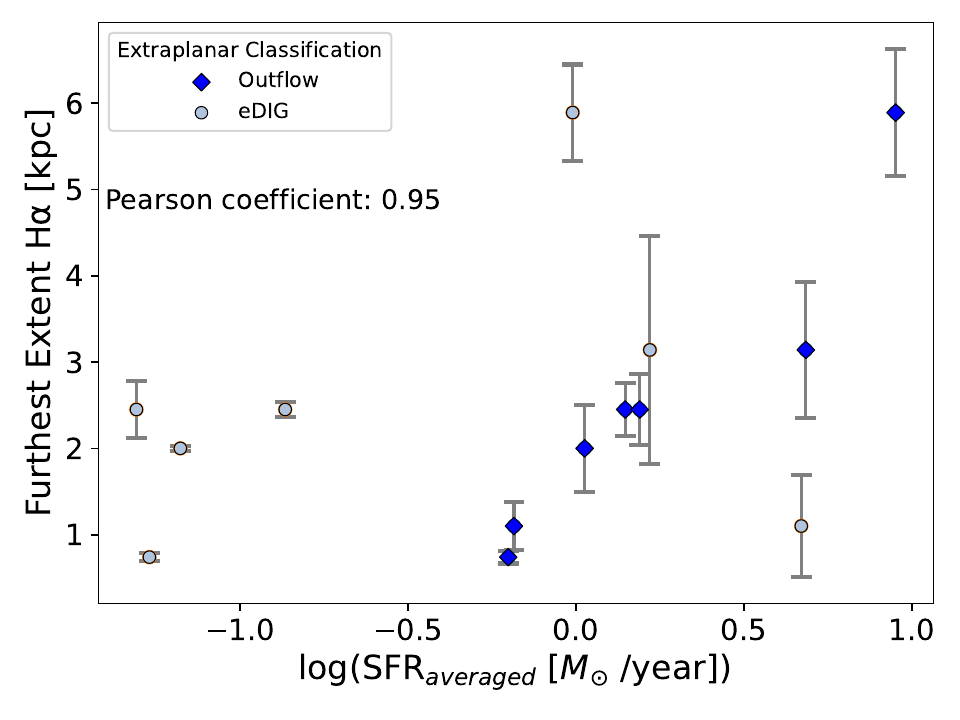}}
  \subfigure{\includegraphics[width=0.5\textwidth, trim=0cm 0cm 0cm 0cm,clip]{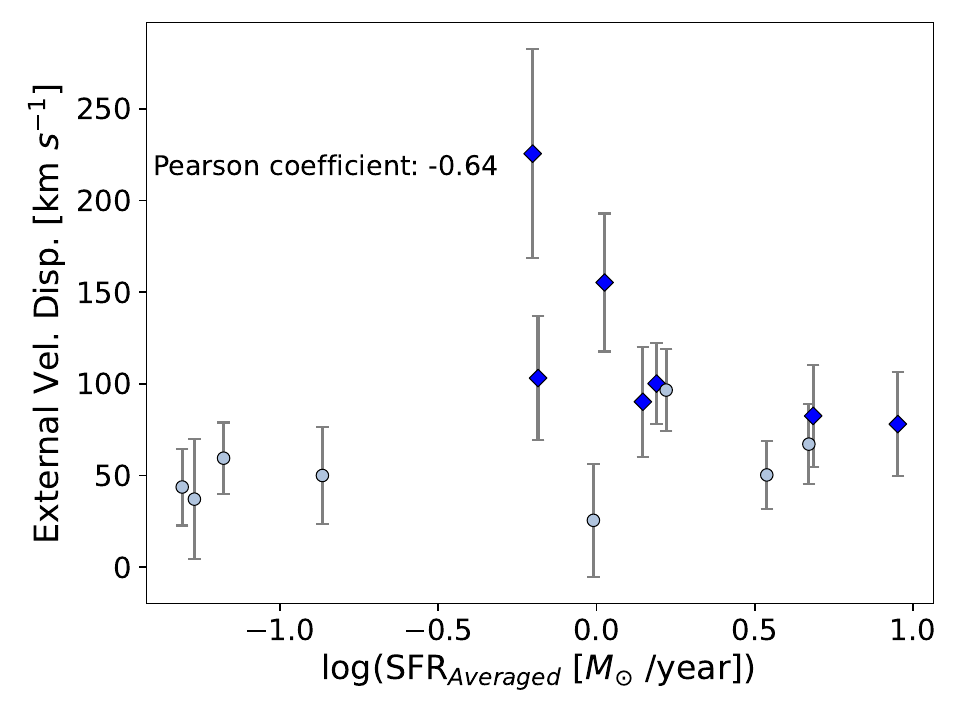}}
  \subfigure{\includegraphics[width=0.5\textwidth, trim=0cm 0cm 0cm 0cm,clip]{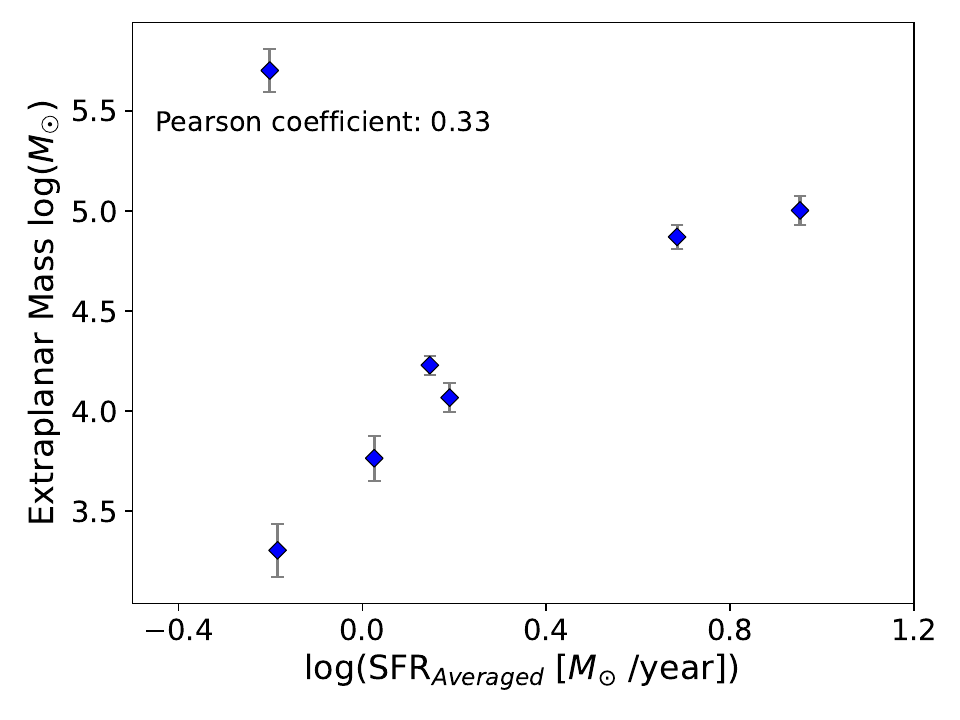}}
  \caption{
  Outflow characteristics for SF SPOGs as a function of log($SFR_{Averaged}$). 
  Top panel: The furthest extent that ionized gas is detected per galaxy.
  Middle panel: The velocity dispersions of the extraplanar gas.
  Bottom panel: Extraplanar gas mass for the SF SPOGs where the gas is labeled as outflowing or weakly outflowing.
  Points are shaped and colored based on if the extraplanar gas is indicative of eDIG (light blue, circle shaped) or outflowing gas (dark blue, diamond shaped).
  Note that eDIG points are plotted in the SF panels for reference, but \textit{not} included in any Pearson correlation tests.
  SFR$_{Averaged}$ is strongly correlated with furthest extent, and perhaps (though affected by one outlier) also with extraplanar mass in SF SPOGs hosting outflows. 
  In contrast, the external velocity dispersion of the gas is actually negatively correlated with SFR$_{Averaged}$ in these systems.
  We find it plausible but not conclusive that the starburst drives these outflows.
  }
\label{SFR Extraplanar Characteristics}
\end{figure*}

SF SPOGs with outflows are examined with the same three properties (furthest extent, velocity dispersion, extraplanar gas mass) as a function of SFR$_{averaged}$ (Figure \ref{SFR Extraplanar Characteristics}).
A clear trend exists between SFR$_{averaged}$ and furthest extent of the ionized gas.
A weak anti-correlation exists between average velocity dispersion of the extraplanar gas and SFR.  
No real trend is seen between SFR$_{averaged}$ and extraplanar gas mass.

We repeat the above comparisons using SFR$_{H\alpha}$, the current SFR measured from H$\alpha$.  Similar, or weaker, correlation coefficient values are found comparing SFR$_{H\alpha}$ to furthest extent, external velocity dispersion, and extraplanar gas mass.  K-S tests comparing the SFR$_{H\alpha}$ distributions of outflow vs non-outflow hosts yield p values of 0.7, only decreasing to 0.56 when including the weak outflowing population with the outflow SF SPOGs.
We conclude that SFR$_{averaged}$ is a stronger indicator of outflow likelihood than SFR$_{H\alpha}$ and that the recent starburst is the main driver of the outflows detected.

For the SF SPOGs, we have similar coefficients to the AGN SPOGs (Figure \ref{SFR Extraplanar Characteristics}).
Taken alone, the correlations with SFR$_{averaged}$ we see here suggests some relationship with the starburst. 
However, these correlations were also present in AGN SPOGs that appear to be driven by AGN. 
While we hypothesize that the recent starburst is the main driver of these outflows, we cannot rule out that AGN feedback could also have played a role in driving them.
There is no evidence of current AGN activity in these SPOGs, but they could be flickering AGN that are turned off (\citealt{1997...AGN...Variability}, \citealt{2015...Schawinski...flickerAGN}).

Mass-loading factors are calculated using SFR$_{averaged}$ are shown in the final column of Table~\ref{mass_galaxy_property_table}.
Values range from 10$^{-6}$ to 10$^{-3}$.
We note that even if our outflow timescales are an order of magnitude too long, the mass-loading factors would still be quite small, and calculating them using SFR$_{H\alpha}$ results in similar mass loading factors $<$1.
We conclude that these outflows are not effective at removing gas from the SF SPOGs, similarly to what was found with AGN and LINER SPOGs.

We note that SF SPOGs may not be properly SPOGs at all.  They were included in the original SPOGs parent sample because their emission line ratios were \textit{consistent} with shocked gas, but it is possible that the emission lines are instead linked to actual current star formation.  If so, this contamination would explain the rarity of starburst-driven outflows in our sample.  A full analysis of the SF SPOGs using similar spectra oriented along the major axis will be carried out in the future to determine which galaxies are truly \textit{post}-starburst.

\subsection{SPOGs as an Evolutionary Stage}

The intention of the SPOGs Survey is to identify galaxies that could be the most recently quenched (or quenching) and identify the physical processes that could be contributing.  By including galaxies containing ionized gas, we are allowing signatures of possible quenching mechanisms (AGN, shocked gas associated with outflows), if they exist.

Our AGN-host SPOGs show a high likelihood of outflows, with high enough mass-loading factors for these AGN-driven outflows to be responsible for the lack of current star formation (intermediate age stellar populations in the SDSS spectra).  We expect this quenching to occur in an inside-out fashion; a full analysis along the major axis of these galaxies could show whether or not star formation has been shut off across the entire galaxy.
This analysis will be the focus of future work done with our LDT SPOGs observations.

The SF-classified SPOGs are a more complicated picture: outflows are rare (or perhaps short-lived) and mass-loading factors are low.  We note that this category is likely contaminated by galaxies that are not truly post-starburst, which may be artificially lowering the outflow rate.  However, we expect from the low mass-loading factors that these outflows are not likely to be able to quench the entire galaxy, and may only be shutting off star formation in the nuclear regions (as probed by SDSS).  

\section{Summary and Conclusions}
We present long-slit spectroscopy from the LDT for a sample of \galaxies{} shocked post-starburst galaxies to search for evidence of outflows. 
To do so, we look for extraplanar ionized gas along the minor axis of each target observed with a signal-to-noise greater than 3 out past the edges of the galactic continuum.
Our conclusions are the following:
\begin{enumerate}
\item We detect extraplanar ionized gas in \extraplanar{} targets (40\%) from our sample. 
AGN and SF galaxies make up the majority of the \ha\ detections (27 out of the \extraplanar{} detections).
AGN make up a majority of the \oiii\ detections (10/15).
\item Using the internal and external velocity dispersions of the SPOGs targets, we can separate the ionized gas into three categories.
9 SPOGs have low external velocity dispersion indicative of eDIG.
4 SPOGs have elevated external velocity dispersions over their internal velocity dispersions but within an error range from which we classify them as weakly outflowing.
18 SPOGs have clear signs of outflows with higher external velocity dispersion than internal velocity dispersion.
This population lies predominantly in the AGN-classified SPOGs sample.
\item Our outflows extend out to around 2 kpc from the galactic center on average, matching what is seen in the literature for outflows.
The furthest extents are seen in both SF and AGN SPOGs with detections out to around 6 kpc.
\item For AGN-dominated galaxies, the extraplanar ionized gas lies in the elevated line ratio areas of the BPT diagram. In contrast, SF and LINER-dominated SPOGs have extraplanar ionized gas with smaller line ratios than their nuclei, consistent with ionization only by HII regions. 
\item \oiii\ luminosity of the AGN does not impact the likelihood of detecting an outflow. However, among AGN hosts with outflows, the current \oiii\ luminosity of the AGN \textit{does} correlate with outflow characteristics like furthest extent, extraplanar velocity dispersion, and total extraplanar gas mass, suggesting the AGN, rather than the past starburst, drove the outflows.

\item There is suggestive evidence that the recent starburst SFR could increase the outflow likelihood for SPOGs with SF-like nuclear line ratios. 
$SFR_{Averaged}$ correlates with the outflows' furthest extent and total extraplanar gas mass (albeit one outlier with gas mass). 
However, this SFR negatively correlates with velocity dispersion.
Further study also needs to be done on the SF population to identify any non-post-starburst star-forming galaxy that might have made it through the selection criteria of SPOGs.

\item We estimate the mass entrained in the extraplanar gas using \ha\ luminosity. 
The total mass in individual outflows ranges from $10^2 M_{\odot}$ to $10^5 M_{\odot}$. 
LINER classified SPOGs in our sample contain less mass in their extraplanar gas than AGN or SF SPOGs do.  These low extraplanar masses may mean that the outflows are simply not removing significant gas from SPOGs, in line with previous findings \citep{2016...Alatalo...2016b...Co...gas, 2022...Smercina...PSBs...gas}.
However, ionized gas mass is likely a lower limit for the multi-phase extraplanar total gas mass.

\item For the AGN SPOGs population, outflows are common.
Similar to SF and LINER SPOGs, AGN SPOGs have mass-loading factors much below 1---indicating that either the galaxies are inefficient at removing gas or we are tracing a small portion of the outflows using ionized gas and a long-slit.
The outflows in AGN SPOGs, however, are likely caused by the central AGN instead of the recent starburst.
\end{enumerate}

\section{Acknowledgements}
We thank the anonymous referee for their constructive feedback, which helped improve this work.

These results made use of the Lowell Discovery Telescope (LDT) at Lowell Observatory.  Lowell is a private, non-profit institution dedicated to astrophysical research and public appreciation of astronomy and operates the LDT in partnership with Boston University, the University of Maryland, the University of Toledo, Northern Arizona University and Yale University. The upgrade of the DeVeny optical spectrograph has been funded by a generous grant from John and Ginger Giovale and by a grant from the Mt. Cuba Astronomical Foundation.

AF, MB, TT, and AMM acknowledge support from the National Science Foundation under grant number 2009416.  MB also acknowledges support through the University of Toledo Research Experiences for Undergraduates program, supported through the National Science Foundation under grant number 1950785.

J.~A.~O.~acknowledges support from the Space Telescope Science Institute Director's Discretionary Research Fund grants D0101.90296 and D0101.90311. Y.~L.~, A.~P.~ and P.~P~acknowledge support from SOFIA grant \#08-0226 (PI: Petric). Y.~L.~acknowledges support from the Space Telescope Science Institute Director's Discretionary Research Fund grant D0101.90281. P.~P~~acknowledges support from NASA grant 80NSSC23K0495. M.~S.~acknowledges support from the William H. Miller III Graduate Fellowship.

Funding for SDSS-III has been provided by the Alfred P. Sloan Foundation, the Participating Institutions, the National Science Foundation, and the U.S. Department of Energy Office of Science. The SDSS-III web site is http://www.sdss3.org/.

\facility{LDT}

\clearpage

\bibliography{ldtmain}{}
\bibliographystyle{aasjournal}

\appendix

\section{Observing Nights}
\label{Observing Appendix}
\begin{table*}[!b]
\begin{tabular}{llllllll}
\hline\hline
SPOG Name & RA & Dec & Redshift & Observing Night & Exposure Time &  Number of Exposures \\
(J2000) & ($^\circ$) & ($^\circ$) &  &  & (seconds) &  (counts) \\
(1) & (2) & (3) & (4) & (5) & (6) & (7)\\
\hline

J0004-0114 & 1.133   & -1.2366 & 0.0888 & 11/28/2021 & 1500 & 3 \\ 
J0015+1411 & 3.9854  & 14.1975 & 0.0834 & 8/14/2020  & 1200 & 3 \\ 
J0034-0017 & 8.5414  & -0.2885 & 0.058  & 9/10/2021  & 900  & 3 \\ 
J0039+0022 & 9.8402  & 0.3722  & 0.0827 & 8/14/2020  & 1200 & 2 \\ 
J0102-0052 & 15.6631 & -0.8788 & 0.0505 & 10/21/2020 & 1200 & 4 \\ 
J0134-0842 & 23.7307 & -8.7108 & 0.0923 & 9/27/2022  & 1200 & 5 \\ 
J0135+1352 & 23.8565 & 13.8831 & 0.0343 & 12/6/2020  & 900  & 3 \\ 
J0141-0015 & 25.2795 & -0.2596 & 0.0562 & 10/13/2021 & 900  & 3 \\ 
J0142+1309 & 25.7014 & 13.1559 & 0.0028 & 12/21/2019 & 900  & 3 \\ 
J0204+0051 & 31.1075 & 0.8648  & 0.0763 & 12/7/2020  & 1100 & 3 \\ 
J0206+1339 & 31.6153 & 13.6646 & 0.0302 & 9/10/2021  & 900  & 3 \\ 
J0237+0042 & 39.4353 & 0.7113  & 0.0809 & 9/27/2022  & 1200 & 3 \\ 
J0316-0002 & 49.2288 & -0.042  & 0.0232 & 9/26/2022  & 900  & 5 \\
J0320-0105 & 50.1451 & -1.0959 & 0.0213 & 12/6/2020  & 900  & 2 \\ 
J0356-0501 & 59.2171 & -5.0299 & 0.0706 & 11/29/2021 & 1200 & 3 \\ 
J0728+3654 & 112.243 & 36.9161 & 0.06   & 10/18/2020 & 1200 & 2 \\ 
 &  &  &   & 10/21/2020 & 1200 & 1 \\ 
J0733+4535 & 113.267 & 45.5836 & 0.0777 & 12/7/2020  & 1100 & 3 \\ 
J0744+4250 & 116.193 & 42.8445 & 0.0508 & 11/28/2021 & 900  & 4 \\ 
J0755+3929 & 118.82  & 39.4889 & 0.0747 & 10/21/2020 & 1200 & 3 \\ 
J0759+5244 & 119.78  & 52.7472 & 0.0192 & 11/29/2021 & 900  & 3 \\ 
J0759+5516 & 119.952 & 55.2801 & 0.0216 & 2/17/2020  & 900  & 3 \\
J0800+3743 & 120.199 & 37.7287 & 0.0416 & 2/16/2020  & 900  & 3 \\ 
J0813+2434 & 123.265 & 24.567  & 0.0204 & 3/15/2020  & 900  & 3 \\ 
J0813+2057 & 123.322 & 20.9659 & 0.0377 & 2/4/2021   & 1000 & 3 \\ 
J0815+3720 & 123.857 & 37.3405 & 0.0397 & 2/4/2021   & 900  & 3 \\ 
J0821+2238 & 125.476 & 22.6413 & 0.0127 & 2/17/2020  & 900  & 3 \\ 
J0824+1823 & 126.145 & 18.394  & 0.0258 & 2/6/2021   & 1000 & 3 \\ 
J0826+4558 & 126.518 & 45.9676 & 0.0071 & 12/21/2019 & 900  & 3 \\ 
J0836+1732 & 129.215 & 17.5355 & 0.0419 & 2/25/2022  & 1200 & 3 \\ 
 &  &  &  & 2/25/2022  & 1500 & 1 \\ 
  &  &  &  & 2/25/2022  & 900 & 1 \\ 
J0843+2205 & 130.92  & 22.0941 & 0.0126 & 2/16/2020  & 900  & 4 \\ 
J0913+2838 & 138.265 & 28.6354 & 0.0225 & 2/26/2022  & 1500 & 3 \\ 
 &  &  &  & 2/26/2022  & 900 & 1 \\

\hline\hline
\end{tabular}
\caption{Table detailing the SPOGs galaxies observed with the DeVeny Spectrograph on the LDT. 
Columns 2 and 3 list the galaxy's right ascension and declination. 
Column 5 lists observing dates.
Column 6 has the exposure time for each data frame taken.
The amount of frames taken are listed in Column 7. 
}
\label{Observing-Table}
\end{table*}

\begin{table*}
\begin{tabular}{llllllll}
\hline\hline
SPOG Name & RA & Dec & Redshift & Observing Night & Exposure Time &  Number of Exposures \\
(J2000) & ($^\circ$) & ($^\circ$) &  &  & (seconds) &  (counts) \\
(1) & (2) & (3) & (4) & (5) & (6) & (7)\\
\hline
J0923+2913 & 140.877 & 29.2318 & 0.0328 & 11/29/2021 & 900  & 3 \\ 
J0937+3335 & 144.268 & 33.5912 & 0.0264 & 11/28/2021 & 900  & 3 \\ 
J0939+1130 & 144.889 & 11.5102 & 0.0197 & 2/16/2020  & 900  & 3 \\ 
J0951+2255 & 147.928 & 22.9319 & 0.0211 & 2/6/2021   & 900  & 3 \\ 
J0956+2029 & 149.183 & 20.486  & 0.0253 & 11/29/2022 & 600  & 6 \\
J1007+3240 & 151.799 & 32.6746 & 0.0287 & 2/6/2021  & 900  & 3 \\ 
J1007+3919 & 151.81  & 39.3299 & 0.0394 & 3/2/2022  & 900  & 3 \\ 
 &   &  &  & 3/2/2022  & 1200  & 1 \\
J1016+1323 & 154.246 & 13.3993 & 0.0325 & 4/30/2022 & 900  & 6 \\
J1025+1647 & 156.413 & 16.7873 & 0.0347 & 2/25/2022 & 900  & 3 \\ 
J1115+2914 & 168.939 & 29.2447 & 0.0465 & 2/25/2022 & 1200 & 3 \\ 
 &  &  &  & 2/25/2022 & 1500 & 1 \\ 
J1115+2615 & 168.968 & 26.2512 & 0.0478 & 2/4/2021  & 1000 & 3 \\ 
J1118+5714 & 169.705 & 57.2498 & 0.0461 & 2/4/2021  & 1000 & 2 \\ 
J1135+2825 & 173.851 & 28.4259 & 0.0337 & 3/2/2022  & 1500 & 3 \\ 
 &  &  &  & 3/2/2022  & 600 & 1 \\ 
 &  &  &  & 3/2/2022  & 900 & 1 \\ 
J1135+1516 & 173.909 & 15.2789 & 0.0363 & 2/26/2022 & 1500 & 5 \\ 
J1136+5811 & 174.11  & 58.1914 & 0.0042 & 3/16/2020 & 900  & 3 \\ 
J1148+5459 & 177.047 & 54.9918 & 0.0081 & 2/17/2020 & 900  & 3 \\ 
J1156+2829 & 179.148 & 28.4904 & 0.0119 & 2/16/2020 & 900  & 3 \\ 
J1207+2702 & 181.952 & 27.0402 & 0.0489 & 3/2/2022  & 1200 & 3 \\ 
 &  &  &  & 3/2/2022  & 1500 & 1 \\ 
 &  &  &  & 3/2/2022  & 900 & 1 \\ 
J1307+5350 & 196.915 & 53.8402 & 0.03   & 3/15/2020 & 900  & 3 \\ 
J1334+3411 & 203.562 & 34.1942 & 0.0236 & 5/27/2020 & 900  & 3 \\
J1358+3901 & 209.505 & 39.0273 & 0.0371 & 4/29/2022 & 900  & 3 \\ 
J1358+0227 & 209.686 & 2.4568  & 0.0241 & 5/27/2020 & 900  & 3 \\ 
J1405+1146 & 211.293 & 11.7714 & 0.0175 & 4/29/2022 & 900  & 3 \\ 
J1411+2531 & 212.761 & 25.5194 & 0.0316 & 2/26/2022 & 900  & 2 \\ 
 &  &  &  & 2/26/2022 & 1500  & 1 \\ 
J1412+3048 & 213.149 & 30.8093 & 0.0144 & 3/2/2022  & 1200 & 3 \\ 
J1430+1607 & 217.635 & 16.1235 & 0.0533 & 4/19/2023 & 900  & 3 \\
J1435+2230 & 218.933 & 22.5059 & 0.0296 & 4/30/2022 & 900  & 3 \\ 
J1439+5303 & 219.826 & 53.0593 & 0.0372 & 5/27/2020 & 900  & 3 \\ 
J1530+5519 & 232.702 & 55.3288 & 0.0462 & 4/29/2022 & 600  & 3 \\ 
J1540+5106 & 235.028 & 51.1137 & 0.0193 & 5/27/2020 & 900  & 3 \\ 
J1543+1659 & 235.754 & 16.9874 & 0.0314 & 5/25/2020 & 900  & 3 \\ 
J1615+3125 & 243.87  & 31.4226 & 0.0584 & 9/10/2021 & 900  & 3 \\
J1624+0936 & 246.209 & 9.6071  & 0.0343 & 4/19/2023 & 900  & 3 \\ 
J1700+2307 & 255.029 & 23.1275 & 0.0568 & 9/26/2022 & 600  & 3 \\ 
 &  &  &  & 9/26/2022 & 1200  & 1 \\
  &  &  &  & 9/26/2022 & 900  & 1 \\

\hline\hline
\end{tabular}
\caption{Continued from Table \ref{Observing-Table}.
} 
\end{table*}

\begin{table*}
\begin{tabular}{llllllll}
\hline\hline
SPOG Name & RA & Dec & Redshift & Observing Night & Exposure Time &  Number of Exposures \\
(J2000) & ($^\circ$) & ($^\circ$) &  &  & (seconds) &  (counts) \\
(1) & (2) & (3) & (4) & (5) & (6) & (7)\\
\hline
J1702+3254 & 255.586 & 32.9125 & 0.0824 & 9/27/2022 & 1200 & 3 \\ 
 &  &  &  & 9/27/2022 & 1200  & 1 \\
  &  &  &  & 9/27/2022 & 900  & 1 \\
J1703+2531 & 255.766 & 25.5297 & 0.0318 & 8/22/2019 & 600  & 3 \\
J1703+6318 & 255.941 & 63.3044 & 0.0801 & 8/14/2020 & 1200 & 3 \\
J1718+3007 & 259.533 & 30.129  & 0.0297 & 8/23/2019 & 600  & 3 \\ 
J1732+5958 & 263.012 & 59.9819 & 0.0293 & 8/22/2019 & 600  & 4 \\
J2129-0010 & 322.337 & -0.174  & 0.03   & 8/22/2019 & 600  & 3 \\ 
J2159-0749 & 329.885 & -7.8237 & 0.0926 & 8/23/2019 & 900  & 3 \\ 
J2225-0004 & 336.378 & -0.0679 & 0.0977 & 8/23/2019  & 1100 & 3 \\ 
J2227-0043 & 336.808 & -0.7333 & 0.0559 & 8/13/2020  & 1100 & 2 \\ 
 &  &  &  & 8/13/2020  & 1200 & 1 \\ 
J2300+0036 & 345.122 & 0.6004  & 0.0818 & 10/21/2020 & 1200 & 3 \\ 
J2313+1528 & 348.281 & 15.4814 & 0.1098 & 10/21/2020 & 1200 & 3 \\ 
J2338-0034 & 354.659 & -0.5726 & 0.0729 & 12/7/2020  & 1100 & 3 \\ 

\hline\hline
\end{tabular}
\caption{Continued from Table \ref{Observing-Table}.
} 
\end{table*}

\end{document}